\documentclass[traditabstract]{aa}
\usepackage{txfonts}
\usepackage{natbib}
\bibpunct{(}{)}{;}{a}{}{,}
\usepackage{graphicx}

\newcommand{\tab} {Table~}
\newcommand{\fig} {Figure~}

\begin{document}
\title{The galaxy stellar mass function and its evolution with time
show no dependence on global environment\thanks{This paper includes data gathered with the 6.5 meter Magellan Telescopes located at Las Campanas Observatory, Chile}\\}

%\subtitle{No dependence of the mass function on global environment}

\author{Benedetta Vulcani\inst{1,2}\thanks{benedetta.vulcani@ipmu.jp}
\and Bianca M. Poggianti\inst{2}
\and August Oemler Jr.\inst{3}
\and Alan Dressler\inst{3}
\and Alfonso Arag\'on-Salamanca\inst{4}
\and Gabriella De Lucia\inst{5}
\and Alessia Moretti\inst{1,2}
\and Mike Gladders\inst{6}
\and Louis Abramson\inst{6}
\and Claire Halliday\inst{7}}
\institute{Astronomical Department, Padova University, Italy,
\and INAF-Astronomical Observatory of Padova, Italy,
\and Observatories of the Carnegie Institution of Science,  Pasadena, CA, USA,
\and School of Physics and Astronomy, University of 
Nottingham, Nottingham NG7 2RD, UK,
\and INAF-Astronomical Observatory of Trieste, Italy,
\and Department of Astronomy \& Astrophysics, University of Chicago, Chicago, IL, USA,
\and INAF-Astronomical Observatory of Arcetri, Firenze, Italy.}

\date{Accepted .... Received ..; in original form ...}

\abstract{We present an analysis of the galaxy stellar mass function in
different environments at intermediate redshift ($0.3\leq z\leq0.8$)
for two mass-limited galaxy samples. We use the IMACS Cluster 
Building Survey (ICBS; $M_{\ast} \geq 10^{10.5} M_{\odot}$) to study cluster,
group and field galaxies at $z=0.3$--$0.45$, and the ESO Distant Cluster Survey 
(EDisCS; $M_{\ast} \geq 10^{10.2} M_{\odot}$) 
to investigate cluster and group
galaxies at $z=0.4$--$0.8$. Our analysis thus includes galaxies 
with masses reaching just below that of the Milky Way.
Excluding the brightest cluster galaxies,
we show that the shape of the mass distribution does not seem to depend on 
global environment,  
Our two main results are: 
(1) Galaxies in the virialised regions of clusters, in groups, and in the field follow similar mass
distributions. 
(2) Comparing the ICBS and EDisCS mass functions to mass functions in the local universe, 
we detect evolution from $z\sim0.4$--$0.6$ to $z\sim0.07$
in the sense that 
the population of low-mass galaxies has grown with time
with respect to the population of massive galaxies.
This evolution is independent of environment, i.e., the same for clusters 
and the field. Furthermore, considering only cluster galaxies, 
we find  that  no differences can be detected in their mass functions 
either within the virialised regions, or when we compare galaxies inside  
and outside the virial radius.
Finally, we find that  red and blue galaxies have different mass
functions. However, the shapes of the mass functions of blue and red galaxies do not seem to depend on 
their environment (clusters groups and the field).}

\keywords{galaxies: clusters: general --- galaxies: evolution --- galaxies: formation --- galaxies:luminosity function, 
mass function}

\titlerunning{No dependence of the mass function on global environment}

\maketitle

\section{Introduction}
In standard $\Lambda$ cold dark matter ($\Lambda$CDM) cosmological models,
cold dark matter haloes form from the gravitational collapse of dark
matter around peaks in the initial density field. Haloes assemble
hierarchically in such a way that smaller haloes merge to form larger and
more massive ones in dense environments \citep{mo96, sheth02}.
According to the current paradigm of galaxy
formation, galaxies form within these haloes due to the cooling of hot
gas. Haloes and galaxies evolve simultaneously and the evolution
of a galaxy is driven by the evolution of its host halo. If the halo is accreted by a
larger halo, the galaxy will also be affected and, in some cases,  
the galaxy's diffuse hot gas reservoir may be stripped, thus removing its
fuel for future star formation (e.g. \citealt{larson80,
balogh00, weinmann06, vandenbosch08}). The evolution can also be 
governed by the interplay between smooth and clumpy cold streams, disk
instability, and bulge formation. Intense and relatively smooth streams can maintain 
an unstable dense gas-rich disk.
Instability with high turbulence and giant clumps is self-regulated by
gravitational interactions within the disk (see, e.g., \citealt{dekel09} and references therein).
Galaxies may also experience major mergers,
which transform late-type galaxies into early-type ones with a
central bulge component (e.g. \citealt{driver06, drory07}). 
Mergers drive gas towards the centres of galaxies, where it can trigger a
burst of star formation and fuel the central black hole, the feedback
from which can heat the remaining gas and eventually quench star
formation (e.g. \citealt{mihos96, wild07, pasquali08,
schawinski09}).

Several studies have shown that there are additional  
external disturbances acting mainly on galaxies in dense environments, and generally 
preventing the survival of
spiral structure.
For example, ram pressure \citep{gunn72, bekki09}
is a gas-driven drag force 
capable of stripping a galaxy of much of its interstellar gas, hence preventing 
the formation of new stars. Galaxy
harassment \citep{moore96,boselli06} is a mechanism that 
strips a galaxy of part of its mass and
may change its morphology as a consequence of frequent
high speed encounters. Harassment has the potential of
changing any internal property of a galaxy within a cluster, including
its gas distribution and content, the orbital distribution of its stars,
and its overall shape.
Finally, cluster tidal forces \citep{byrd90}
can act with different efficiency
depending on the environment in such a way that  
gas-rich field galaxies infalling into larger
structures may be transformed into gas-poor lenticular galaxies.

For all these reasons, galaxies are expected to be strongly influenced 
by the environment in which they reside during their evolution. 
It is also broadly accepted that the evolution of a galaxy 
depends very strongly on its stellar mass. 
\cite{kauffmann03} found that colour, specific star formation rate, and internal
structure are strongly correlated with galaxy stellar mass.
\cite{pasquali09} demonstrated
that the star formation and AGN activity of galaxies have 
a much stronger dependence on stellar mass than on halo mass. 
\cite{thomas10} argued that the formation of
early-type galaxies is environment-independent and driven only
by self-regulation processes and intrinsic galaxy properties such as
mass.

Distinguishing the separate contribution of environmental processes and 
processes driven by intrinsic galaxy properties is clearly critical to understanding galaxy evolution.
In the `nature versus nurture' debate, mass represents the primary
``intrinsic property'' expected to be closely related to the initial conditions.
The `environment', on the other hand, represents all the possible external processes 
that can influence galaxies in their evolution.

There are many papers in the literature where the effect on galaxies of 
the global environment and mass are 
analysed separately (see e.g. \citealt{guo09,iovino10,mercurio10,peng10}).
The basic idea is to study variations in galaxy properties as a function
of mass fixing the environment, or as a function of the environment fixing the galaxy mass. 
However, studies of galaxy properties as a function 
of mass often ignore the possibility that the galaxy mass distribution itself may vary with environment.

In this paper we investigate whether (and how) the effects of mass and environment are related
and whether the environment can influence the galaxy masses themselves. In particular, 
we will consider the galaxy stellar mass distribution in different environments. 
Galaxy stellar masses can be estimated reasonably easily, even though the uncertainties are 
still relatively large, but there is not a unique way to describe the environment.

Different methods of estimating stellar masses agree reasonably well within the errors. Examples 
include \citet{bj01},  who use a relation between the stellar mass-to-light ratio and the galaxy colour, 
and  \citet{bolzonella10}, who use a SED fitting technique based on the code {\tt Hyperzmass}, a modified version
of the photometric redshift code {\tt Hyperz} \citep{bolzonella00}. 
Note, however, that different choices of the IMF, stellar population synthesis model, SFR history, metallicity 
and extinction law can systematically affect the stellar mass estimates (see, e.g., 
\citealt{marchesini09, muzzin09, conroy10}).

When describing or quantifying the environment, 
it is possible to refer to either global or local environments. 
As discussed in detail in \cite{muldrew11}, 
there is no universal environmental measure and the most suitable method generally depends on
the spatial scale(s) being probed.
Concerning the global environment, galaxies are commonly subdivided into, for instance,  
superclusters, clusters, groups, the field and voids, which can be expected to roughly 
correspond to the galaxies' host halo mass. The local 
environment is generally described using estimates of the local density, which can be calculated following several definitions and methods.

Several works have focused on the galaxy mass distribution and its evolution in 
the field, but very little is known about the galaxy stellar mass function in clusters.
Studies focused on field galaxies 
(\citealt{drory05, gwyn05, fontana06, bundy06, pozzetti07, drory09}; and \citealt{baldry12}), 
present the mass function for 
the global field galaxy population at different redshifts 
and obtain consistent results. \cite{fontana04, fontana06, bundy06, borch06}; and \cite{pozzetti10} 
demonstrated that galaxies with $M_{\ast} \geq 10^{11}M_{\odot}$
exhibit relatively modest evolution in their total mass function from $z=1$ to $z=0$. 
This implies 
that the evolution of relatively-massive objects (with masses
close to the local characteristic mass)  is essentially complete
by $z\sim1$. On the other hand, the mass functions of less massive 
galaxies were found to evolve 
more strongly than those of massive ones, displaying a rapid rise since $z\sim1$. 
Readers interested in the evolution of the mass function at $z>1$ should see 
\cite{elsner08, kaj09, marchesini09, caputi11, gonzalez11} and \cite{mortlock11}.

There have been a number of mass function studies separating galaxies by morphology. 
Galaxies with different morphologies and star formation histories can contribute in
different ways to the mass function.
Broadly speaking, galaxies can be separated into different populations such as early and late types 
according to their star formation histories (using rest-frame colours, SEDs or spectra)
or their structure (using quantitative structural parameters or visual 
morphologies).  
\cite{balogh01} analysed the environmental dependence of the luminosity function
and the associated stellar mass function of passive and star forming galaxies in the
Two Micron All Sky Survey. They found that 
the mass function of field star-forming galaxies has a much steeper high mass end than that of field passive galaxies.
In clusters, however,  both star-forming and passive galaxies have a steep high mass end.

Subdividing field galaxies according to their colour in the local universe  
\citep{baldry04, baldry06, baldry08} and at intermediate redshifts \citep{borch06, bolzonella10}
it has been found that the mass function is bimodal, with 
early-type galaxies dominating at high masses and late-types
mostly contributing at intermediate and low masses.
Moreover, these authors also find that the mass functions 
of early- and late-type galaxies evolve differently with redshift.

\cite{morph} studied for the first time
the mass function of cluster galaxies, finding 
quite a strong evolution  with redshift. 
Clusters in the local universe
are proportionally more populated by low mass galaxies than clusters at high $z$. 
This study concluded, first, that the mass growth caused by star formation plays a crucial
role in driving the observed evolution; second, that this star-formation-driven mass growth 
must be accompanied by infall of
galaxies onto clusters; and third, that the mass distribution of 
the infalling galaxies may be different from that of cluster galaxies. 
To test this last result, a comparison was made between the mass function of
cluster galaxies and that of field galaxies found in the literature.
Preliminary analysis suggests that  at high masses ($\log M_{\ast} / M_{\odot}
\geq 11$), the mass functions of
field and cluster galaxies at high-$z$ seem to have rather similar shapes.
However, the situation at intermediate/low
masses is not clear since different field studies
give quite different results in this mass range. 
Using the field mass function of \cite{ilbert10}, it seems as if 
field galaxies have a steeper mass function than cluster galaxies at intermediate/low masses,
thus suggesting significant environmental mass
segregation.  In contrast, the results of 
Bundy (2005)\footnote{These data are the combination of \cite{bundy05} and
\cite{bundy06} (Bundy, private communication).} suggest 
that there are no large
differences between the mass distributions of galaxies in different environments at high $z$.
Based on these results, it remains unfortunately unclear whether 
field and cluster galaxies have similar or different mass
distributions. 
It is also important to note that the preliminary results presented in \cite{morph} were obtained 
using inhomogeneous data and slightly different redshift ranges, so 
definite conclusions cannot be drawn.

On the theoretical side,
\cite{moster10} found 
a correlation between the stellar mass of the central galaxy
and the mass of the dark matter halo.
Using N-body simulations, they found that the 
clustering properties of galaxies are predominantly
driven by the clustering of the halos and subhaloes in which they reside, and
provided a model to predict clustering as a function of stellar 
mass at any redshift. 
This result could also suggest that the {\it total} (central $+$ satellites) mass function
may depend on environment. However, the correlation between the total 
galaxy stellar mass function and the mass of the 
parent halo has not yet been studied. It would be very interesting
to understand whether
simulations predict a mass segregation with environment, considering 
the initial and evolved halo
mass and how they predict the evolution with redshift 
in different environments. 
This would allow us to understand the role of the halo mass
in influencing the evolution of galaxy masses.
This analysis is deferred to a forthcoming paper (Vulcani et al., in preparation).

\section{Aims of this work}\label{aim}

So far, observational studies have shown that 
in different environments early- and late-type galaxies are present in different proportions and
follow different mass distributions.
As mentioned above, the theoretical expectation is that the stellar mass 
of central galaxies depends on the environment.
This may also raise the expectation that 
the total mass function could be different in different environments.
The main goal of our work is to test this  
possibility by studying the stellar mass distribution as a function
of the halo mass. Note that to compare mass functions of galaxies in clusters, groups and the 
field means also to compare mass functions of galaxies hosted in haloes with different masses.

We want to investigate whether the mass
function is ``universal'' and, if this is the case, how this came about. 
The main questions of this paper are:
(i) do observations suggest that the mass function at intermediate redshifts is
driven by the halo mass? (ii)
does the mass function of red and blue galaxies separately depend on halo mass?
(iii)  does the evolution of the mass functions depend on global environment (a proxy for 
halo mass)? Alternatively, we wish to know whether 
 it is possible for the galaxy mass distribution to be unaffected by where 
the galaxies form and which halo they grow in. If that is the case, 
we will need to ask whether there is some mechanism imprinting the same stellar mass
distribution on all galaxies,
regardless where they are.

The analysis presented in this work is complementary to that presented in \cite{ld}.
There we
analysed the role of the local density in shaping the mass function, using a nearest-neighbour-based density
measure largely independent of dark matter halo mass \citep{muldrew11}.
In that paper, we addressed the following questions:
does the mass function depend on local density at low- and intermediate-$z$, 
both in clusters and in the field? How does the mass function change with
local density?
These two papers therefore address different points. The two ways to define
the ``environment'' are not equivalent and provide different
information (the effect of halo masses versus local phenomena).  
Sometimes the differences between local and global environment 
are subtle and confusing, and it is 
therefore possible that part of the results presented in this paper will surprise the reader. 
They are not in line with what it is generally expected.
We will show here that, even if the mass function does depend on local density 
(as we show in \citealt{ld}), the differences in local density distributions in clusters
compared to the field are insufficient to induce a difference in the mass functions in these 
global environments.  Thus, the investigation of the dependence of the mass functions on 
global environment that we carry out in this paper is an independent test of whether the global 
environment alone is able to produce a difference in the galaxy mass function.
Note that \cite{ld} brings together the results of both studies, 
contrasting the role of global and local environments in shaping the mass function.

The study of the mass functions of galaxies has only been developed in the last years, while
much more effort has been devoted to characterising luminosity functions. 
The literature contains  a variety of papers that carefully analyse 
luminosity functions, their evolution with redshift and their dependence
on environment. At first sight, it seems reasonable to expect that the stellar mass function 
will simply mirror the luminosity 
function behaviour, given that stellar masses are derived from luminosities. For this reason
it is commonly thought that results derived from the luminosity function would also apply to the mass function.
However, this is not necessarily the case. 
A simple linear correlation between mass and luminosity does not exist:
not all galaxies have the same mass-to-light ratio. In cases such as  passively-evolving old
galaxies the mass-to-light ratios are fairly constant, but galaxies with different star formation histories 
can have significantly different mass-to-light ratios. As a consequence, the luminosity function 
does not provide, in general, direct
information on the mass function (see Appendix A). It is therefore not unreasonable to expect that the
environmental effects on the luminosity and mass functions may not be the same.  

Furthermore, it is important to note that the samples used when analysing
luminosity and mass functions should be assembled following different criteria.
Traditionally, luminosity functions are studied in magnitude-limited samples, 
where a cut in luminosity is performed. In contrast, mass function studies 
require, ideally, mass-limited samples, including all galaxies more massive than 
a given limit regardless of their colour or morphological type. 
As discuss in Appendix B of \cite{morph}, the choice of a magnitude limit implies 
a natural mass limit below which the sample is incomplete.
Hence the mass distribution derived from a magnitude-limited sample is
meaningless because it is affected by incompleteness: galaxies will be missing
at masses below the limit corresponding to the mass of a galaxy with the reddest colour and the
faintest magnitudes in the sample. It is therefore clear that mass-limited samples are needed 
in environmental and evolutionary studies of the mass function,

The goal of this paper is thus to compare the galaxy stellar
mass distribution in cluster regions, cluster infalling regions, groups,
and the field using homogeneous data. This will allow us to establish 
whether and by how much the total galaxy stellar mass function depends
on global environment at a fixed redshift. Note that it is interesting to consider the mass function
not only in clusters but also in their outskirts, where galaxies will have time to become part
of the clusters themselves by $z=0$. We will also divide galaxies by colour, 
separating the mass functions of star-forming and passive galaxies in all environments. 

The paper is organized as follows. In 
section~\ref{dataset} we present the
two surveys we use in our analysis, the IMACS Cluster 
Building Survey (ICBS) and the ESO Distant Cluster Survey (EDisCS). 
In section~\ref{def_env} we define
the different environments. 
Section~\ref{res} shows the results, with 
section~\ref{mf_envi} presenting the galaxy
stellar mass function as a function of the global environment, 
section~\ref{comp} compares
our findings with some results from the literature, quantifying the evolution
of the mass function in different environments 
We then analyse the galaxy
stellar mass function in clusters (\S\ref{mf_cluster}), dividing it 
by colour (\S\ref{mf_colour}). In section~\ref{disc} we discuss our results,
explaining their
implications for the evolution of mass
functions (\S\ref{ev}), and the dependence of the mass distribution on galaxy properties (\S\ref{rb}).
We then contrast the different roles played by the global and local environments in shaping the mass function (\S\ref{gl_loc}). 
Finally, in section~\ref{conc} we summarise our results.

Throughout this paper, we assume $H_{0}=70 \, \rm km \, s^{-1} \,
Mpc^{-1}$, $\Omega_{m}=0.30$, and $\Omega_{\Lambda} =0.70$.  The adopted
initial mass function (IMF) is that of \cite{kr01} in the mass range 0.1--100
$M_{\odot}$. All magnitudes used in this paper are Vega magnitudes.

\section{Data set} \label{dataset}
In this paper, we take advantage of two different surveys to carry out an
analysis of the mass function.  We use ICBS data to characterise
galaxies at intermediate redshifts ($0.3\leq z\leq 0.45$) and EDisCS data to study a
large sample of galaxies at $0.4\leq z \leq 0.8$.

The ICBS provides homogeneous spectroscopic data of galaxies in
several environments. Spectroscopy yields accurate redshifts, and therefore the
membership to the different environments is well established.

EDisCS contains a much larger sample of cluster and
group galaxies, although spectroscopic redshifts are available for only a
subset of them. Photometric redshifts are therefore used, even though they are
less reliable. In Appendix A of \cite{morph}, we showed  that
the galaxy mass function determined using photo-$z$'s and photo-$z$
membership  agrees with the mass function
determined using only spectroscopic members and spectroscopic
completeness weights in the mass range across which they overlap.

\subsection{ICBS} \label{dataicbs}
The ICBS (Oemler et al. 2012a,  submitted)
is focused on the study of
galaxy evolution and infall onto clusters from 
a clustercentric radius  $R\sim5\,$Mpc to the cluster inner cores.
Data have been acquired using the wide field of the 
Inamori-Magellan Areal Camera
and Spectrograph (IMACS) on the Magellan-Baade telescope  
for four fields containing clusters at intermediate redshift.

The ICBS aims at defining a homogeneous sample of clusters by  
selecting the most massive cluster per comoving volume at any  
redshift. Clusters were selected using the Red-Sequence Cluster Survey  
method \cite{gladders00}, either from the RCS  
itself, or from the Sloan Digital Sky Survey in regions of the sky not  
covered by the RCS. 
As described in Oemler et al.\ (2012a, submitted), direct imaging in the $griz$
bands was obtained for two of the fields with the f/2
camera of IMACS. Imaging in the $BVRI$ bands was obtained for the other two 
fields using the Wide Field CCD camera on the du Pont
Telescope. In addition, very deep $r$-band photometry, complete to 
$r = 25.0$, was obtained for all fields with IMACS. Spectroscopic targets
were selected from $r$-band photometry down to a limiting
magnitude $r=22.5$ for all fields.

The IMACS f/2 spectra have an observed-frame resolution of $10\,$\AA{}
full width at half-maximum with a typical $S/N\sim 20$--$30$ per resolution element in the
continuum. 
In each $28^{\prime}$ diameter IMACS field, 
spectra for
65\% of the galaxies brighter than $r \sim22.5$ were taken 
on the 6.5m Baade Telescope at Las Campanas.
Of those  
observed, only about 20\% failed to yield redshifts, or turned out to  
be stars. Details of the data and its analysis are presented in Oemler et al.\  
(2012a, submitted) and Oemler  et al. (2012b, submitted).

The data discussed in this paper come from four fields  containing
rich galaxy clusters at $z = 0.33$, $0.38$, $0.42$ and 0.43, as well as other 
structures at different redshifts. 
In this paper, we restrict our analysis to ICBS galaxies in the redshift range
$0.3<z<0.45$ in all environments. This was done in order to
focus on a limited redshift range so that a common magnitude and mass limit could be set up to $z=0.45$.
Consequently, we have analysed data for galaxies in these four rich clusters
together with field and group galaxies within the chosen redshift range in all four fields.
In \tab~\ref{tab:ic_cl} we provide information for the four target clusters. 
Velocity dispersions ($\sigma$) 
were calculated using ROSTAT \citep{beers90}. Galaxies were accepted as cluster members if 
their redshifts placed them within $\pm3 \sigma$ from the cluster redshift.

\begin{table*}
\centering
\begin{tabular}{lccccc}
\hline
Cluster name & $z$ & $\sigma$ & $R_{200}$	&$N_{\rm gals}$ & $N_{\mbox{\rm gals above} }$\\
	   &	& (km$\,$s$^{-1}$)		& Mpc	&&the mass limit\\
\hline
SDSS0845A& 0.3308 &$975\pm53$ &2.03	&181&100\\
RCS1102B & 0.3857 &$695\pm33$ &1.40	&208&96\\
SDSS1500A & 0.4191 &$528\pm37$ & 1.04	&81&50\\
RCS0221A& 0.4317 &$798\pm43$ &1.57	&201&111\\
\hline
\end{tabular}
\caption{List of ICBS clusters analysed in this paper, with cluster
name, redshift, velocity dispersion, $R_{200}$ and number of cluster 
member galaxies ($\pm 3 \sigma$ from cluster redshift). 
\label{tab:ic_cl}}
\end{table*}

\begin{figure*}
\centering
\includegraphics[scale=0.3, angle=-90]{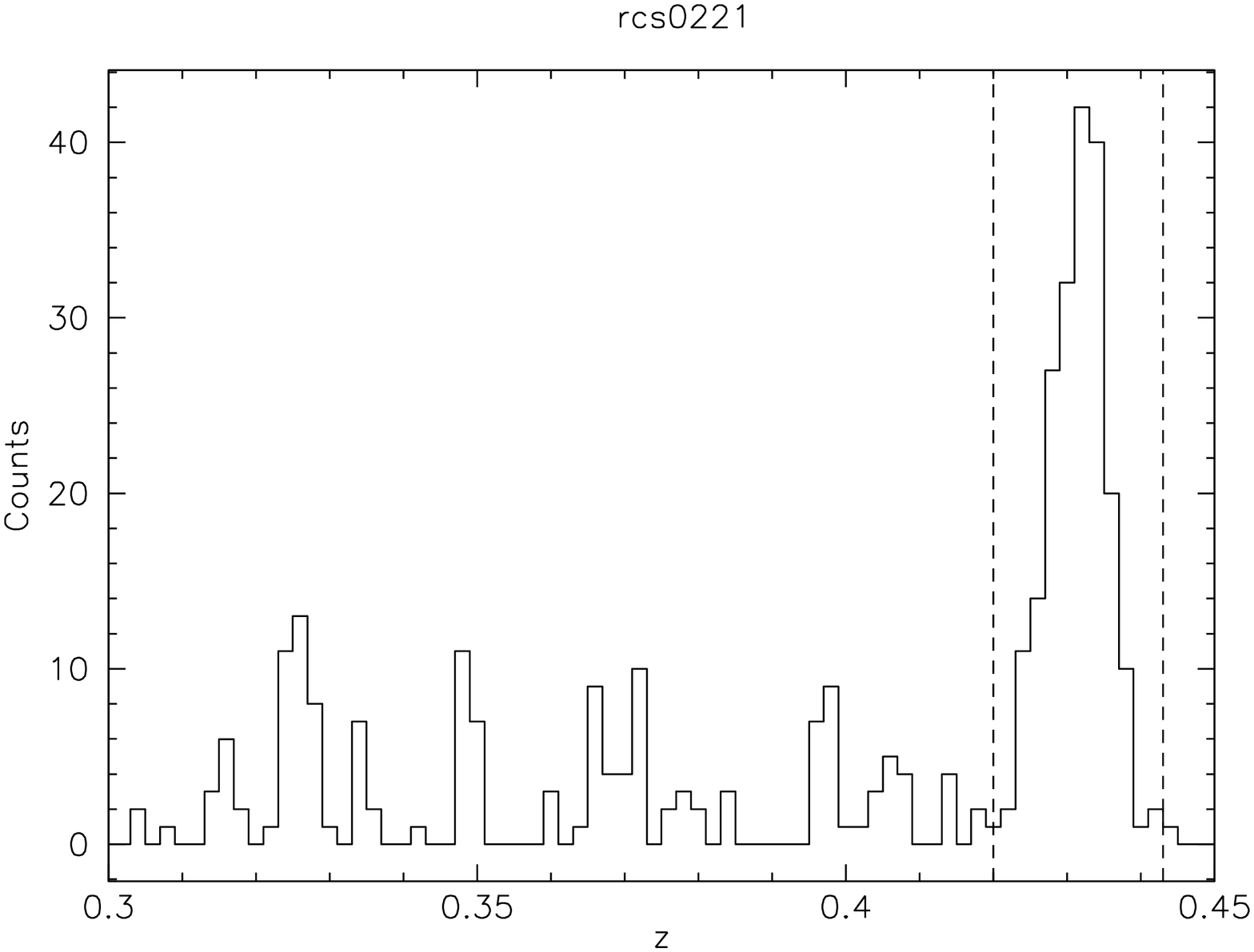}
\includegraphics[scale=0.3, angle=-90]{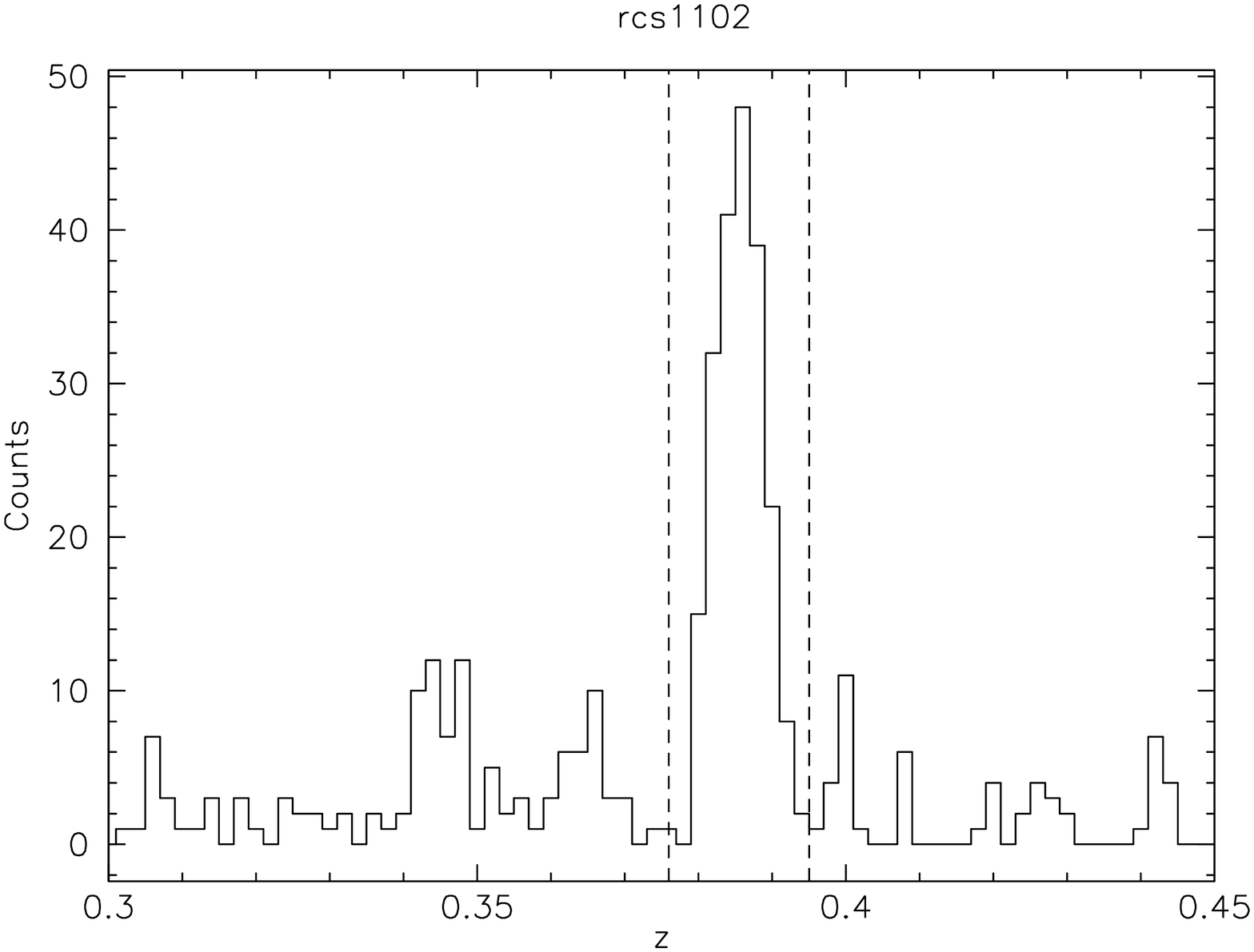}
\includegraphics[scale=0.3, angle=-90]{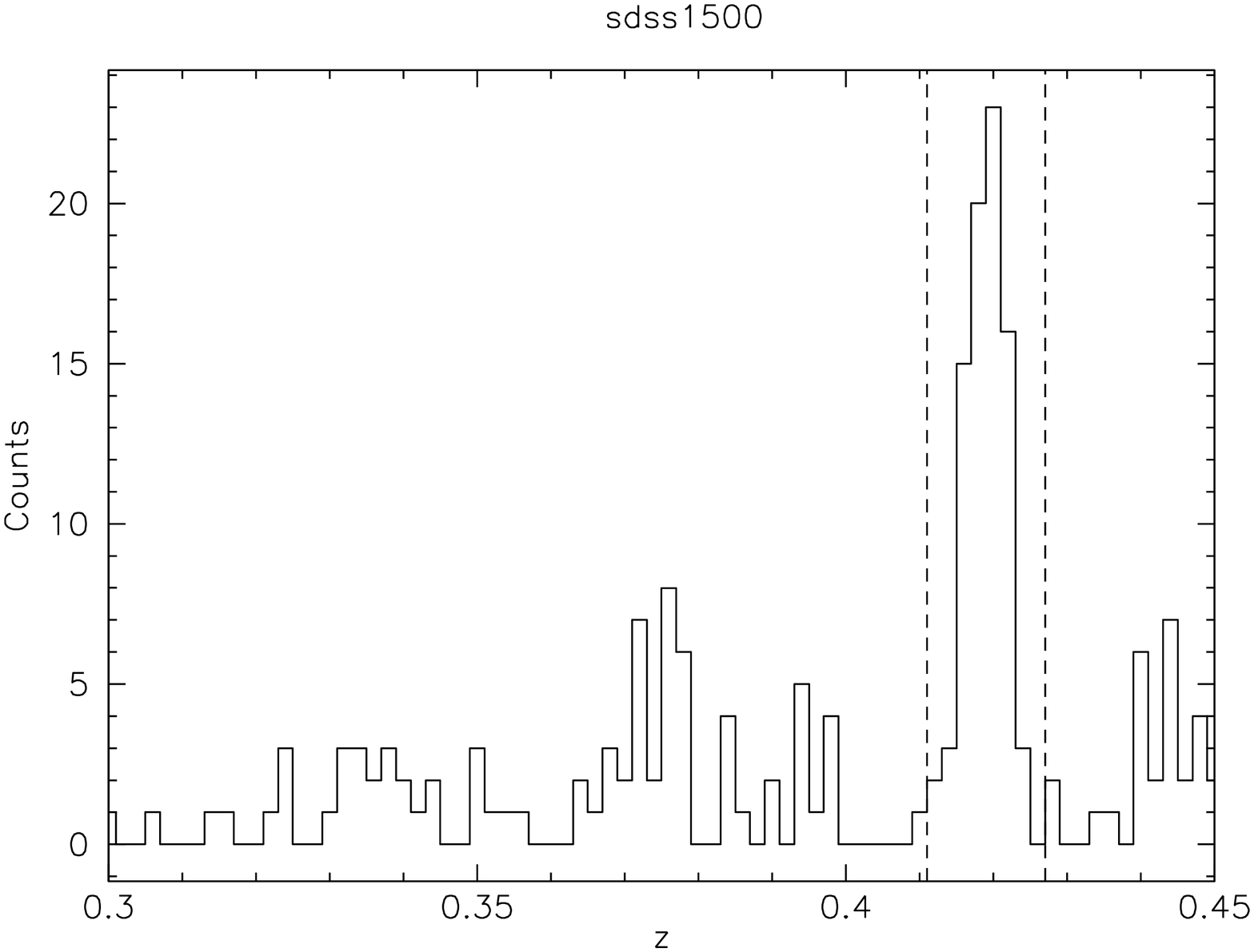}
\includegraphics[scale=0.3, angle=-90]{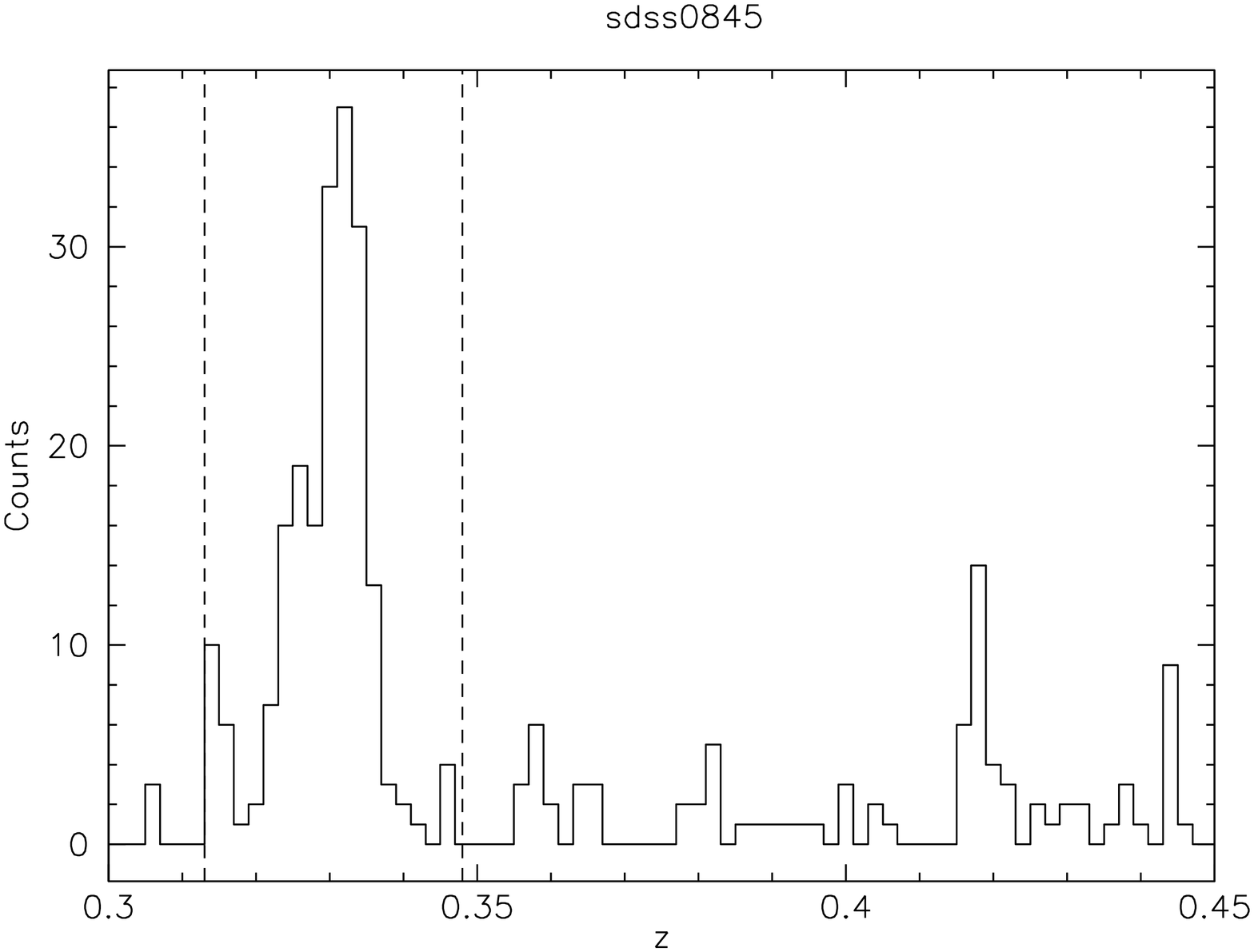}
\caption{ICBS: redshift distribution in the four fields observed by the survey: 
RCS0221, RCS1102, SDSS0845, and  SDSS1500. The cluster regions 
($\pm 3 \sigma$ from cluster redshift) are also indicated (vertical dotted lines).
\label{z_histo}}
\end{figure*}

Since the projected density of cluster/supercluster members
is low at the large clustercentric distances probed by the ICBS, 
our sample necessarily includes $\sim1000$ ``field'' galaxies at redshifts $0.2 < z < 0.8$ in each
survey field. This gives us the opportunity to compare the evolution of cluster
and field galaxies over this redshift range. 
\fig~\ref{z_histo} shows the redshift distribution
for the four fields {\it RCS 1102}, {\it RCS 0221}, {\it SDSS 1500}, and
{\it SDSS 0845} in the redshift range considered. The cluster regions ($\pm 3 \sigma$ from the cluster redshift) 
are also indicated.

Absolute magnitudes were determined using INTERREST
\citep{taylor09} from the observed photometry. 
The tool interpolates the rest-frame magnitudes from the observed photometry
in bracketing bands (see \citealt{rudnick03}). The code 
uses a number of template spectra to carry out this interpolation. 
Rest-frame colours are derived from the interpolated rest-frame apparent magnitudes.

Galaxy stellar masses are derived using the relation 
between  $M/L_{B}$  and rest-frame $(B-V)$ colour given in \cite{bj01}
\begin{equation}
\log_{10}(M/L_{B})=a_{B}+b_{B}(B-V).
\end{equation}. For a Bruzual \& Charlot model with solar metallicity and a \cite{salpeter55} IMF 
(0.1-125 $M_{\odot}$), $a_B=-0.51$ and $b_B= 1.45$. 
Since our broadband photometry does not cover the entire field of   
the redshift survey, galaxies without the required photometry had   
synthetic colours calculated from the flux-calibrated IMACS spectra.
The error in the measured masses
is $\sim$0.3 dex.
All our masses are scaled to a \cite{kr01} IMF, by adding -0.19 dex 
to the logarithmic value of the Salpeter masses.

The magnitude completeness limit of the ICBS spectroscopy 
is $r \sim 22.5$. At the redshift limit of the ICBS sample, $z \sim 0.45$, 
we determine the value of the
mass of a galaxy with an absolute $B$ magnitude corresponding to $r=22.5$,
and a rest-frame colour $(B-V) \sim 1$, the reddest colour 
of galaxies in ICBS clusters.
From this we determine that the ICBS mass completeness limit at the redshifts of interest
is $M_{\ast} = 10^{10.5} M_{\odot}$.
The colour distributions of both the ICBS and EDisCS samples (see \fig\ref{UB} and \fig\ref{UB2}) 
confirm that they are complete and unbiased down to our mass limit.

In this paper, galaxies are given weights proportional to the inverse of the
spectroscopic incompleteness.  Since the main galaxy property that  we wish to
analyse is galaxy stellar mass, we compute the
incompleteness correction taking into account the number of
galaxies for which an estimate of the mass is available. 
Above our mass limit, 91\% of all galaxies brighter than $r=22.5$ with 
available spectroscopy have mass 
estimates.\footnote{The remaining 9\% of the relevant galaxies have no mass estimates due to bad photometry or colours.}

The incompleteness correction
depends on both the apparent magnitude and the position in the field. 
We subdivided each field into three different regions
according to 
their distance from the centre
of the main cluster ($R/R_{200}\leq 1; 1<R/R_{200}\leq 2; {\rm and}\ R/R_{200}>2$)\footnote{$R_{200}$ 
is defined as the radius delimiting a 
sphere with interior mean density 200 times the critical 
density of the universe at that redshift,
and is commonly used as an approximation of the cluster 
virial radius. The $R_{200}$  values for our structures are computed 
from the velocity dispersions using the formula 
$$
R_{200}=1.73\frac{ \sigma}{1000 ({\rm km \, s}^{-1})}\frac{1}{\sqrt{\Omega_{\Lambda}+\Omega_{0}(1+z)^{3}}}h^{-1}  \quad  ({\rm Mpc})
$$ 
}
and we then determined the completeness weights in 0.4 $r$ 
magnitude bins around each galaxy as
the ratio of the number of galaxies with a spectroscopic redshift and a given mass
to the number of galaxies in the original photometric
catalog.

For all clusters we excluded the Brightest Cluster Galaxy (BCG), 
identified as the most luminous cluster member, because its characteristics
could alter the general conclusions.
The final mass-limited ICBS sample 
consists of 596 galaxies with 
$M_{\ast} \geq 10^{10.5} M_{\odot}$ in all environments. After correcting for incompleteness, this number becomes 1295.

\subsection{EDisCS} \label{dataediscs}
The EDisCS multiwavelength photometric and spectroscopic survey of galaxies
\citep{white05} was developed to characterise both
the clusters themselves and the galaxies within them.  It observed 20
fields containing galaxy clusters at $0.4< z <1$.
These clusters were selected from the Las Campanas Distant Cluster Survey
(LCDCS) catalog \citep{gonzalez01}.

For all 20 fields deep optical multiband
photometry obtained with FORS2/VLT  \citep{white05} and near-IR photometry obtained 
with SOFI/NTT is available.
ACS/HST mosaic $F814W$ imaging was also acquired for 10 of the highest redshift
clusters \citep{desai07}.
FORS2/VLT was additionally used to obtain spectroscopy for 18 of the fields
\citep{halliday04, milvang08}. 

The FORS2 field covers $R_{200}$ for all
clusters with the exception of {\it cl 1232.5-1250}, for which the field only
covers $0.5 R_{200}$ \citep{poggianti06}. The $R_{200}$
values were computed from the velocity dispersions
by \cite{poggianti08}.

Photometric redshifts were computed for each object 
from both optical and infrared  data using 
two independent codes, a modified version of the 
publicly-available Hyperz \citep{bolzonella00} and the code of
\cite{rudnick01}, as modified in
\cite{rudnick03} and \cite{rudnick09}. The accuracy of both methods is $\sigma (\delta z)
\sim 0.05$--$0.06$, where $\delta z = \frac{z_{\rm spec}-z_{\rm phot}}{1+z_{\rm spec}}.$
Photo-$z$ membership (see also \citealt{delucia04} and \citealt{delucia07} for details)
was determined using a modified version of the technique
first developed in \cite{brunner00}. The probability $P(z)$ for a
galaxy to be at redshift $z$  is integrated in a slice
$\Delta z = \pm 0.1$ around
the cluster redshift to give $P_{\rm clust}$ for the two codes. 
A galaxy was rejected as a member if $P_{\rm clust}$ is smaller than a certain probability
$P_{\rm thresh}$ for either code.  For each cluster $P_{\rm thresh}$ was 
calibrated using the spectroscopic redshifts to maximise the 
efficiency of spectroscopic
non-member rejection while retaining at least $\sim 90\%$ of the confirmed
cluster members independently of rest-frame $(B-V)$ or
observed $(V-I)$. 

For EDisCS galaxies, stellar masses were estimated  following
\cite{bj01} and then scaled to a \cite{kr01} IMF.  
Total absolute magnitudes were derived from the photo-$z$ fits
\citep{pello09} and rest-frame luminosities calculated using the methods of 
\cite{rudnick03} and \cite{rudnick06}, as presented in \cite{rudnick09}.
As a check, stellar masses for spectroscopic members were also estimated 
using the {\tt kcorrect} tool 
\citep{blanton07}\footnote{http://cosmo.nyu.edu/mb144/kcorrect/}. These masses agree with 
the photo-$z$ based ones within the errors. A detailed discussion on stellar mass estimates and
the consistency between different methods can be found in \cite{morph}.

\begin{table}
\centering
\begin{tabular}{lccc}
\hline
name & z & $\sigma$ & $R_{200}$\\
	   &	& (${\rm km \, s}^{-1}$)		& ${\rm Mpc}$\\
\hline
Clusters		&	&			& \\							
Cl 1232.5$-$1250 &  0.5414  &1080$^{+119}_{-89}$&1.99 \\
Cl 1216.8$-$1201 &  0.7943  &1018 $^{+73}_{-77}$&1.61\\ 
Cl 1138.2$-$1133 &  0.4796  &732 $^{ +72}_{-76}$ &1.41\\	  
Cl 1411.1$-$1148 &  0.5195  &  710$^{+125}_{-133   }$& 1.32\\ 
Cl 1301.7$-$1139 &  0.4828  &  687$^{+81}_{-86     }$&1.30\\  
Cl 1353.0$-$1137 &  0.5882  &  666$^{+136}_{-139   }$ &1.19\\ 
Cl 1354.2$-$1230 &  0.7620  &  648$^{+105}_{-110   }$& 1.08\\ 
Cl 1054.4$-$1146 &  0.6972  &  589$^{+78}_{-70     }$&0.99\\  
Cl 1227.9$-$1138 &  0.6357  &  574$^{+72}_{-75     }$ &1.00\\ 
Cl 1202.7$-$1224  & 0.4240  &  518$^{+92}_{-104    }$&1.07\\  
Cl 1059.2$-$1253  & 0.4564  &  510$^{+52}_{-56     }$ &1.00\\ 
Cl 1054.7$-$1245	& 0.7498  &  504$^{+113}_{-65    }$& 0.82\\ 
Cl 1018.8$-$1211	& 0.4734  &  486$^{+59}_{-63     }$&0.91\\  
Cl 1040.7$-$1155	& 0.7043  & 418$^{+55}_{-46     }$& 0.70\\
\hline
Groups		&	&			& \\		  	 	   	
Cl 1037.9$-$1243	& 0.5783  & 319$^{+53}_{-52     }$ &\\
Cl 1103.7$-$1245b	& 0.7031  & 252$^{+65}_{-85     }$&\\ 
Cl 1103.7$-$1245a	& 0.6261  & 336$^{+36}_{-40     }$ &\\
Cl 1420.3$-$1236	& 0.4962  & 218$^{+43}_{-50     }$& \\
Cl 1119.3$-$1129	& 0.5500  & 166$^{+27}_{-29     }$ &\\

\hline
\end{tabular}
\caption{List of EDisCS clusters and groups analysed in this paper, with cluster
name, redshift, velocity dispersion and (only for clusters) $R_{200}$ (from \citealt{halliday04,
milvang08,poggianti08}).
\label{tab:ed_cl}}
\end{table}

To build the EDisCS mass-limited sample, we used all photo-$z$ members of all clusters and groups.
Using the photo-$z$ membership selection instead of the 
spectroscopic one we maximise sample size while still retaining adequate 
quality. Furthermore, the spectroscopic magnitude limit ($I=22$ or $I=23$, depending on redshift)
would correspond a stellar mass limit  $M_{\ast} =10^{10.6} M_{\odot}$
\citep{vulcani10}. Using photo-$z$'s allows us to push the mass limit to significantly 
lower values. Our adopted conservative magnitude completeness limit for the EDisCS photometry is $I \sim
24$ (the sample remains close to 90\% complete at $I\sim25$, White et al.\ 2005).
For the most distant cluster in our sample ({\it cl 1216.8-1201} at $z\sim0.8$) the
stellar mass of a galaxy with an absolute $B$ magnitude corresponding to $I=24$
and $(B-V) \sim 0.9$, the reddest colour, is $M_{\ast} = 10^{10.2} M_{\odot}$.
We take this value as the EDisCS stellar mass completeness limit for the photo-z
sample. As before, BCGs were excluded.
Table \ref{tab:ed_cl} presents the list of clusters used and some
basic properties.

The final mass-limited EDisCS sample of galaxies with 
$M_{\ast} \geq 10^{10.2} M_{\odot}$
consists of 2962 objects.

\section{Definition of the different environments}\label{def_env}
The mass-limited galaxy samples defined above were divided into different 
environments. 
For the ICBS survey, the cluster galaxy sample 
includes all the cluster members, i.e., galaxies lying within 
$3\sigma$ of the cluster redshift. These are then subdivided 
according to their
clustercentric distance. Galaxies in the ``{\it cluster virialised regions}'' 
are defined as those with $R/R_{200}\leq1$, and those with $R/R_{200}>1$
are said to live in the 
``{\it cluster outskirts}''. Moreover,
we further subdivide the cluster core (virialised region)  into three zones: inner 
parts ($R/R_{200}\leq0.2$), intermediate parts 
($0.2<R/R_{200}\leq0.6$), and outer parts ($0.6<R/R_{200}\leq1$).

The group galaxy sample was defined from a group 
catalog constructed using the standard method of \cite{hg82}. 
We identify groups by a friends-of-friends  
technique where the linking velocity distance used to connect friends  
is $350\,{\rm km\, s}^{-1}$ and the projected linking length 
$D_L$ scales with the incompleteness of the data as
$$
D_L = D_0 \left[ I(r,\alpha,\delta) \frac{\int_{-\infty}^{M_{\rm pair}}  
\Phi(M) dM}{\int_{-\infty}^{M_{\rm lim}} \Phi(M) dM} \right]^{-1/2}. 
$$In this formula, $D_0 = 0.40\,{\rm Mpc}$ is the linking length at a fiducial redshift
$z_{\rm fid} = 0.30$. $I(r,\alpha,\delta)$ is the incompleteness of the data  
at a given $r$ magnitude and position in the field, as described in  
Oemler et al. (2012a, submitted). The numerator is the integral of the galaxy luminosity function  
to the limiting absolute magnitude at the distance of the galaxy pair,  
corrected for galaxy evolution as described in Oemler et al. (2012b, submitted) and the  
denominator is the integral of the galaxy luminosity function to the  
absolute magnitude limit at the fiducial redshift.

Although this method makes efficient use of all the data, it produces groups
whose properties vary systematically with  
redshift because of the definition of $D_L$. However, since we 
only consider a fairly narrow redshift range, this drawback does not affect our analysis. 

Finally, we call ``field galaxies'' all those galaxies that are not cluster members.

Within the EDisCS sample we are able to reliably identify only two main environments: 
(1) clusters, with their virialised regions, outskirts,  
and the inner, intermediate and outer regions defined as before, and  
(2) groups. EDisCS clusters are defined as systems with
velocity dispersion $\sigma >400\,{\rm km \, s}^{-1}$. Cluster members
are selected using the photo-$z$ technique described above. 
Groups are defined as systems 
with at least eight spectroscopic members and 
velocity dispersions in the range $150\, {\rm km \, s}^{-1} \leq \sigma \leq 
400\,{\rm km  \, s}^{-1}$. Group members are also selected
using the photo-$z$ method. \tab~\ref{summary}
summarises the definitions adopted to characterise the different environments,
separately for ICBS (upper panel) and EDisCS (bottom panel).

\begin{table*}
\centering
\begin{tabular}{lll}
\hline
\multicolumn{2}{l}{ICBS}		& \multicolumn{1}{l}{definition}\\
\hline
\multicolumn{2}{l}{     cluster virialised regions}	&within $3\sigma_{\rm cluster}$, $R/R_{200}\leq 1$\\
	& cluster inner part 			& within $3\sigma_{\rm cluster}$, $R/R_{200}\leq0.2$\\
	& cluster intermediate part 		&within $3\sigma_{\rm cluster}$, $0.2<R/R_{200}\leq0.6$\\
	& cluster outer part 			&within $3\sigma_{\rm cluster}$,  $0.6<R/R_{200}\leq1$\\
\multicolumn{2}{l}{cluster outskirts} 	&within $3\sigma_{\rm cluster}$, $R/R_{200}> 1$\\
\multicolumn{2}{l}{groups	}	& Geller-Huchra group finding  method \\
\multicolumn{2}{l}{field	}	& galaxies not belonging to clusters\\
\hline
\multicolumn{2}{l}{EDisCS} & \multicolumn{1}{l}{definition}\\
\hline
\multicolumn{2}{l}{     cluster virialised regions}		& $\sigma_{\rm struct} >400\,{\rm km\, s}^{-1}$, $R/R_{200}\leq1$, photo-$z$ membership\\
	& cluster inner part 			& $\sigma_{\rm struct} >400\,{\rm km\, s}^{-1}$, $R/R_{200}\leq0.2$, photo-$z$ membership\\
	& cluster intermediate part 		& $\sigma_{\rm struct} >400\,{\rm km\, s}^{-1}$, $0.2<R/R_{200}\leq0.6$, photo-$z$ membership\\
	& cluster outer part 			& $\sigma_{\rm struct} >400\,{\rm km\, s}^{-1}$, $0.6<R/R_{200}\leq1$, photo-$z$ membership\\
\multicolumn{2}{l}{cluster outskirts}	& $\sigma_{\rm struct} > 400\,{\rm km\, s}^{-1}$, $R/R_{200}>1$, photo-$z$ membership\\ 
\multicolumn{2}{l}{groups	}		& $150\,{\rm km \, s}^{-1} \leq \sigma_{\rm struct} \leq 400\,{\rm km \, s}^{-1}$, photo-$z$ membership\\
\hline
\end{tabular}
\caption{Definitions adopted to characterise the different environments for ICBS (upper panel) and EDisCS (bottom panel).\label{summary}}
\end{table*}

Note that the definitions of groups are not the same in both samples. However, 
this does not affect our findings since we never compare them directly. 

As mentioned above, we exclude 
the BCGs from both samples when we consider the cluster regions because they could alter the general trends. 
The properties of BCGs are in many aspects
very different from those of other galaxies, and they the subject
of many specific studies (e.g., \citealt{aragon98,whiley08,fasano10}).

Above the mass limit of $M_{\ast}\geq 10^{10.5}$ the ICBS sample contains 
178 cluster galaxies in the virialised regions,
177 in the  cluster outskirts, and 241 field galaxies, 90 of which belong to groups 
(see \tab\ref{num} for completeness-corrected numbers). 
The EDisCS sample has 268 galaxies in the cluster virialised regions, 749 in the outskirts,  
and 620 group galaxies above a mass limit of $M_{\ast}\geq 10^{10.2}$ (see \tab\ref{num}).

\begin{table*}
\centering
\begin{tabular}{lcc}
\hline
ICBS		& $N_{\rm obs}$ & $N_{\rm weight}$ \\
       &	($M_{\ast}\geq 10^{10.5}\,M_{\odot}$)                 & ($M_{\ast}\geq 10^{10.5}\,M_{\odot}$) \\
\hline
cluster virialised regions&	178 	& 339\\
cluster outskirts 	&	177	& 374\\
groups		&	90	&199\\
field		&	241	&581\\
\hline
EDisCS &	$N_{\rm obs}$ & $N_{\rm obs}$  \\
       &	($M_{\ast}\geq 10^{10.2}\,M_{\odot}$)                 & ($M_{\ast}\geq 10^{10.5}\,M_{\odot}$) \\
\hline
cluster virialised regions	&1268	& 842\\
cluster outskirts		&749	&484 \\
groups			&620  	& 421\\
\hline
\end{tabular}
\caption{Number of galaxies in each environment above the stellar mass completeness limits.  
$N_{\rm obs}$ refers to the observed number of galaxies, 
while  $N_{\rm weight}$ refers to the incompleteness-corrected numbers obtained taking 
into account the completeness weights described in the text. Upper panel: ICBS sample.  lower panel: EDisCS sample.  
For EDisCS we show sample sizes for two mass limits, $M_{\ast}\geq 10^{10.2}\,M_{\odot}$, the intrinsic mass limit of the survey, 
and $M_{\ast}\geq 10^{10.5}\,M_{\odot}$, the mass limit of the ICBS sample for comparison. 
}
\label{num}
\end{table*}

\section{Results}\label{res}

Using the mass-selected samples defined above we have built 
histograms characterising the
mass distribution of galaxies located in different environments 
The width of each mass bin is $0.2\,$dex. 
For the ICBS sample each galaxy is weighted by
its spectroscopic incompleteness correction. Note that we do not need 
to apply this correction for EDisCS galaxies since we are using photo-$z$ defined samples.  
The errorbars on the $x$-axis represent the width of the bins. The errorbars on the
$y$-axis are computed combining the uncertainties derived from Poisson errors \citep{gehrels86}
and those resulting from cosmic variance.\footnote{For field samples, we computed 
the errors given by cosmic variance by estimating the differences between the field mass function and the mass functions
obtained using each field separately. Similarly, for clusters we computed  the errors due to cluster to cluster variance
by comparing the mass functions of each cluster separately.}
Throughout this paper  
all mass functions are normalised using the total integrated stellar mass above the mass completeness limit \citep{rudnick09}. 
so that the total galaxy stellar  mass in each histogram is equal to 1.
Such a normalisation allows us to focus our analysis on the {\it shape} of the mass functions and not on the number density, 
which is obviously very different across the different environments.

To quantify the differences between different mass functions, we
perform Kolmogorov-Smirnov (K--S) tests which tell us whether we can disprove
the null hypothesis that two data sets are drawn from the same parent distribution. Since
the standard
K--S test does not consider (in)completeness when compiling the
cumulative distributions (i.e., it assigns weight 1 to each object), 
we modified it in such a way that the relative importance
of each galaxy in the cumulative distribution depends on its weight.
Obviously, in the case of photo-$z$ samples (EDsiCS) all galaxies have weight$\,=1$ so
using the modified test is equivalent to using the standard one.
For ICBS data we use the modified K--S test to take into account the spectroscopic incompleteness.  

We recall that a ``positive'' (statistically significant) K--S result
provides robust proof that the  two distributions are different, but a negative
K--S result does not mean that the distributions are identical. 
Therefore, visual inspections of the mass distributions, in particular at their high-mass ends,
could be useful. 

In the analysis that follows we always use the mass limit of $M_{\ast}
\geq 10^{10.5} M_{\odot}$ for the ICBS survey, while for EDisCS we use its own
mass limit ($M_{\ast} \geq 10^{10.2} M_{\odot}$) for internal comparisons but the 
ICBS mass cut when comparing both surveys.

\subsection{The mass function in different environments is very similar} \label{mf_envi}
First of all, we wish to characterise the galaxy stellar mass distribution of
galaxies located in different global environments, to see whether it
depends on the region in which galaxies reside.
\begin{figure*}
\centering
\includegraphics[scale=0.4]{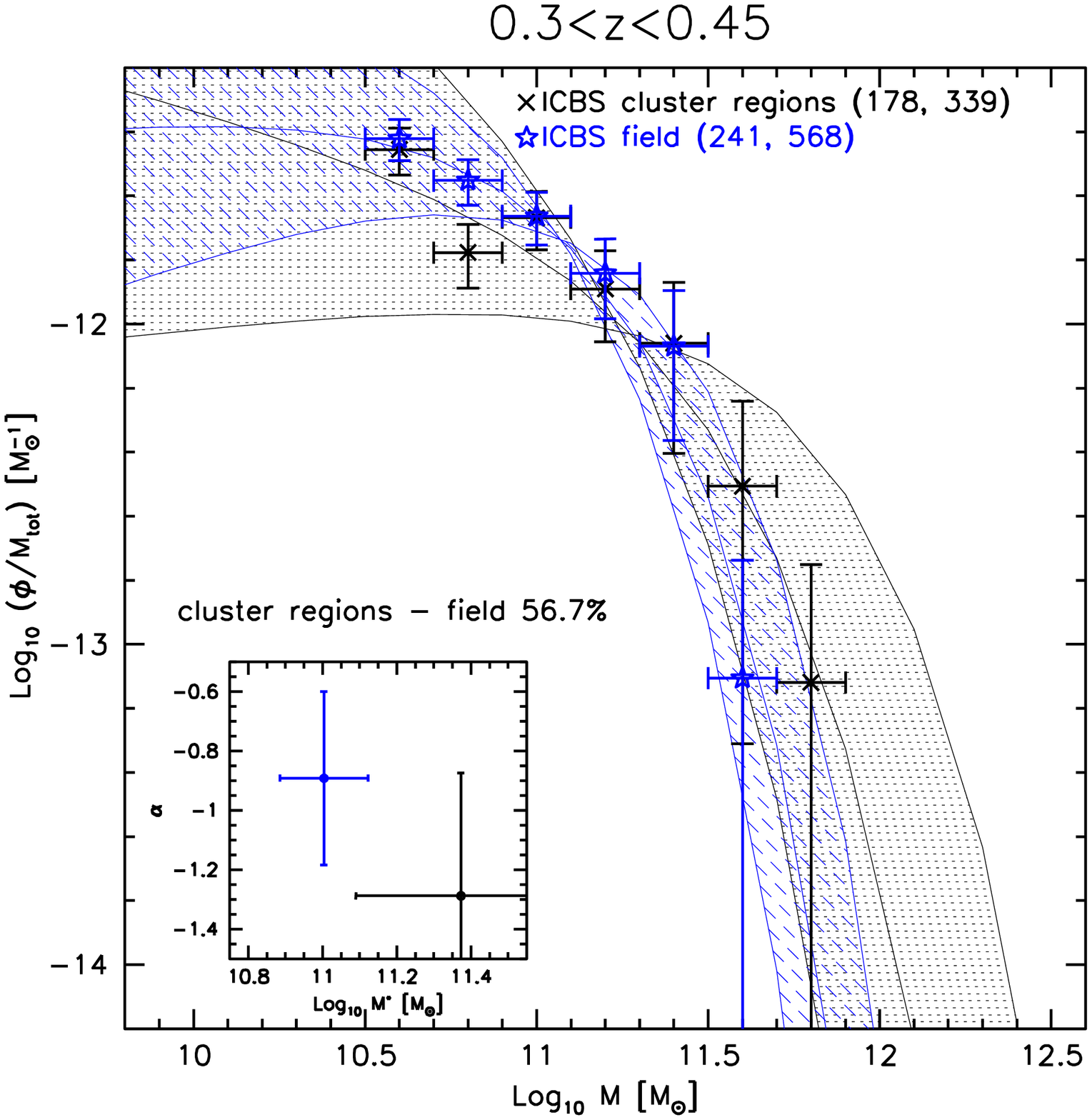}
\includegraphics[scale=0.4]{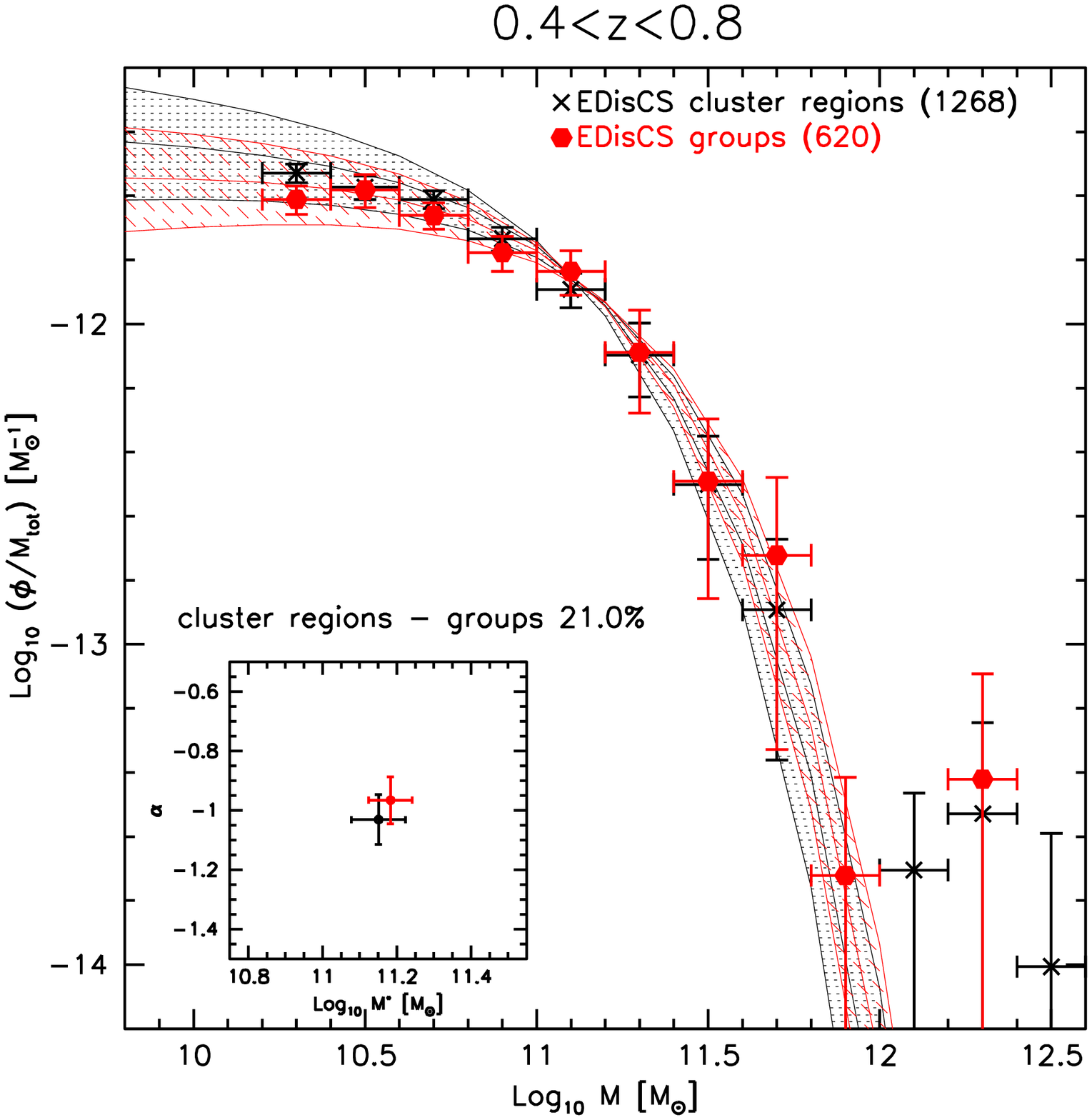}
\caption{Observed mass functions and \citet{schechter76} function fits for galaxies 
in the different environments. Left panel: ICBS cluster regions (black crosses and solid  line) and 
field (blue empty stars and dotted line). Right panel: EDisCS clusters (black crosses and solid line) and groups (red filled hexagons and dotted line). 
Mass functions are normalised  using  the total integrated stellar mass above the mass completeness limits. 
Errorbars on the $x$-axis represent the width of the bin. Errorbars on the
$y$-axis are computed combining the  Poissonian errors \citep{gehrels86}  
and the uncertainties due to cosmic variance and cluster-to-cluster variations. 
Shaded area represent $1\sigma$ errors on the Schechter fits.
The labels at the top show both the observed 
and completeness-weighted galaxy numbers for ICBS, while for EDisCS only observed numbers are given.  
The K--S probabilities are also shown as percentages.  
The bottom left in-sets in each panel show the $M^*$ and $\alpha$ Schechter fit parameters 
with 1$\sigma$ errors. No significant differences are evident between the galaxy stellar mass distributions in different environments. 
For ICBS, similar results are also obtained comparing the cluster regions and
the non-cluster regions (plot not shown).
\label{mf_env2}}
\end{figure*}

To begin, we compare galaxies in the cluster virialised regions ($R/R_{200}\leq 1$) with
field and group galaxies. Using ICBS data only
we can compare the most widely different environments, 
the cluster virialised regions and the field. The left panel of \fig~\ref{mf_env2} shows no 
significant differences between the shapes of the
galaxy stellar mass distributions for these two environments. 
The K--S test is unable to reject the null hypothesis that  the 
two distributions are drawn from the same sample ($P_{\rm K--S}\sim 46\%$).
A visual inspection of the mass functions provides confirmation of the K--S result:  their shapes 
are similar, within the errors, in both environments.
To increase the quality of the data statistics, we also compare the
galaxies in the 
cluster virialised regions with those in the field+cluster outskirts. 
Again, we do not find detectable differences ($P_{\rm K--S}\sim 40\%$; plot not shown). 

In the right panel of \fig~\ref{mf_env2} we compare EDisCS cluster and group galaxies 
reaching lower masses than before ($M_{\ast} \sim 10^{10.2} M_{\odot}$).  
Yet again,  there is no clear evidence of a 
dependence of the  shape of the mass function on the environment. 
The  K--S test is still inconclusive  ($P_{\rm K--S}\sim 21\%$) even with the better statistics 
provided by the larger samples. A visual inspection clearly reveals that the shapes of the mass functions are very similar. 

Photometric redshifts can be uncertain,  
especially for blue galaxies (see \citealt{rudnick09}).  
In principle, photo-$z$ errors could bias the stellar mass functions derived for EDisCS galaxies. 
However, we checked, first, that the magnitudes at which \cite{rudnick09} found discrepancies 
in the galaxy luminosity functions correspond to lower stellar masses than those considered here 
(our $M_{g}$ is always brighter than $-20.2$). In addition, the photo-$z$ counts are fully consistent with 
the statistically background-subtracted counts for both red and blue galaxies. Furthermore, for the mass range in common, 
the mass function determined from photo-$z$'s is in agreement (within the
errors) with the mass function determined using only spectroscopic members after taking into account spectroscopic completeness \citep{morph}. These tests provide reasonable confidence in the results obtained using photo-$z$ techniques.

For both ICBS and EDisCS, the results concerning the low mass end of the mass functions 
are statistically robust. At the high mass end, however, the size of the uncertainties do not allow a reliable  
characterisation of the mass functions and their putative differences. It is therefore 
not possible to conclude whether the apparent similarities are real or simply due to the size of the errors.

The strength of our results can be double checked by considering analytical 
fits to the mass functions using a least-square-fitting method. 
Assuming that  the number density $\Phi(M)$ of galaxies with stellar mass $M_\ast$
can be described by a \cite{schechter76} function, the galaxy stellar mass function can
be written as
\begin{equation}\label{eq:sc}
\Phi (M) = (\ln10) \, \Phi^*  10^{(M-M^*)(1+\alpha)} \, \exp\left(-10^{(M-M^*)}\right),
\end{equation}
where $M = \log (M_\ast/M_\odot)$, $\alpha$ is the low-mass-end slope,
$\Phi^*$ the normalisation and $M^*  = \log (M_\ast^* /M_\odot)$ corresponds 
to the characteristic stellar mass 
$M_\ast^*$ at which the mass function exhibits a rapid change in slope. 
Schechter function fits consider only galaxies above our conservative mass completeness limits.
\tab\ref{tab:fit_global} gives the best-fit Schechter function parameters for the mass functions of 
galaxies in different environments (see also \S\ref{comp} and \S\ref{mf_cluster}).

\begin{table*}[!t]
\caption{Best-fit Schechter function parameters ($M_\ast^{*}$, $\alpha$,  $\Phi^{*}$) for the mass functions of galaxies in different environments and redshifts (see \S\ref{comp} and \ref{mf_cluster}). Note that because the mass functions 
have been normalised as described in the text, the values of $\Phi^{*}$ do not provide meaningful information, but they
are given here for completeness. \label{tab:fit_global}}

\centering
\begin{tabular}{llccr}
\hline
	&	&	$\log (M_\ast^*/M_\odot) $ & 	$\alpha$ &$\log \Phi^{*} $\\
\hline	
{\bf ICBS} &	cluster regions	&$11.37\pm0.28$ 	& $-1.29\pm0.41$ 	&	$31\pm28$\\
	&	cluster outskirts 	&$11.40\pm0.11$	& $-1.29\pm0.00$	&	$33\pm7$\\
	&	field		&$11.00\pm0.12$	&$-0.89\pm0.29$	&      	$134\pm40$\\
\hline	
{\bf EDisCS} &   	cluster regions 		&$11.15\pm0.07$ 	& $-1.03\pm0.08$ 	&	$143\pm24$\\
	&			cluster outskirts		&$11.25\pm0.06$  &	$-1.03\pm0.00$	&	$74\pm5$ \\
	&			groups			&$11.18\pm0.06$ &	$-0.97\pm0.08$	&	$71\pm10$  \\
\hline
{\bf WINGS} &   	cluster regions	&$10.82\pm0.13$  &	 $-0.88\pm0.31$  	& $219\pm75$ 	\\
\hline	
{\bf PM2GC} &   	general field	& $10.96\pm0.06$ &	$-1.12\pm0.12$ 	&	 $174\pm32$ 	\\
\hline
\end{tabular}
\end{table*}

Note that because the mass functions 
have been normalised, 
the values of $\Phi^{*}$ do not provide meaningful information on the true galaxy density. Since 
we are only concerned about comparing the shapes of the mass functions, in what follows 
we only carry out comparisons involving $M^{*}$ and $\alpha$.  

Allowing all the parameters to be fitted freely, we find that the mass functions 
in different environments show comparable $\alpha$ and $M^*$. Explicitly, 
given that the errors in these two parameters are correlated, 
we explored a grid of $\alpha$ and $M^*$ and calculated the corresponding 
$\chi^2$ values and from these the likelihood of two mass functions having the same pair of 
parameters. We found that that the fitted mass functions agree within $1\sigma$. 
This exercise gives additional support to our finding that the  shapes of the mass functions 
do not seem to depend on the global environment. 
This result is particularly robust for EDisCS since the better statistics allow for 
good constraints on the Schechter function 
parameters.\footnote{For EDisCS, we also computed the best-fit \citet{schechter76} parameters using the STY \citep{sandage79} method (see, e.g, \citealt{marchesini09}) which is an unbinned maximum likelihood method. In this case the parameters 
are compatible within $2\sigma$. We do not adopt this method throughout the entire paper 
because it is not trivial to take into account the ICBS completeness weights.}

It is important to point out that, in agreement with previous studies  
(e.g., \citealt{bell03, baldry08}), we find that
a Schechter function is not able to adequately describe 
the highest mass end of the cluster and group mass functions. 
A sort of ``bump'' or excess is seen at   
$M_{\ast} \sim 10^{12.3} M_{\odot}$ in both clusters and groups.  
However, a  Schechter function fits well the distributions for  masses below $M_{\ast} \sim 10^{12} M_{\odot}$. 
It could be argued that for EDisCS this bump may be 
caused by contamination by interlopers since we use photo-$z$'s to determine 
membership. However, most of the high-mass photo-$z$ members are also 
spectroscopic members, and galaxies in the bump do not show larger photo-$z$ 
errors than lower mass galaxies. We note that a similar bump has also been detected in the 
mass function of the low-$z$ spectroscopic cluster galaxy sample in the WINGS survey
(WIde-field Nearby Galaxy-cluster Survey, \citealt{fasano06}. See  \S\ref{comp}).

In the  ICBS survey we have also considered more narrowly-defined environments separately.
When comparing the mass functions of galaxies in clusters, groups and the field outside groups we 
fail to detect any obvious differences in their shapes.  
Even though the statistical uncertainty is larger and the mass functions are thus noisier, 
the $P_{\rm K--S}$ is always inconclusive. 

Confidence in the robustness of our results is reinforced by 
the fact that the mass functions derived from EDisCS and ICBS independently show a similar lack of environmental
dependence. These two surveys have different strengths. 
ICBS provides spectroscopically-defined relatively small samples with
very reliable separation into the different environments. On the other hand, EDisCS provides a larger 
galaxy sample extending to lower masses, albeit affected by the intrinsic photo-$z$ uncertainties.

These results seem be at odds with the findings of \cite{kovac10}. 
They studied $\sim 8500$ galaxies from the zCOSMOS-bright 
redshift survey in the COSMOS field. They
found that  the shape of the stellar mass function is different 
for group, field and isolated galaxies  up to $z\sim 0.7$ at least. 
Their stellar mass function shows an upturn at low masses in the group environment.
They also found that more massive galaxies preferentially reside
in groups.  However, their samples reach much lower masses 
than ours (down to $M_\star \sim 10^{9.5} M_\odot$) and, more 
importantly, their selection of group and field galaxies are very different
from the one adopted here. Therefore, a direct comparison of the results
cannot be made at this point.

We finish this section by noticing that, even though 
the mass functions 
appear to have similar shapes in all the environments studied here, there is a hint that
a small difference may be present. 
The mass functions seem to extend to different
maximum masses, the so-called mass function cut-off.
In the ICBS survey the most massive galaxies in clusters have
$M_{\ast} \sim 10^{11.9} M_{\odot}$, while 
in the field they have 
$M_{\ast} \sim 10^{11.7} M_{\odot}$, and in groups $M_{\ast} \sim 10^{11.6} M_{\odot}$.
In EDisCS, the virialised regions of clusters seem to 
contain galaxies as massive as $M_{\ast} \sim 10^{12.5} M_{\odot}$, while cluster outskirts
only reach $M_{\ast} \sim 10^{12.1} M_{\odot}$ and
groups $M_{\ast} \sim 10^{12.3} M_{\odot}$.

\subsection{The evolution of the mass functions is very similar in different environments}\label{comp}

We have shown in this paper that for masses $M_{\ast} \geq 10^{10.2}-10^{-10.5} M_{\odot}$
the shape of the galaxy stellar mass function at intermediate redshifts
does not seem to depend on the global environment in which galaxies reside. 
We have carried out a complementary study in the local universe (Calvi et al., in preparation) 
using a mass-limited sample ($M_{\ast} \geq 10^{10.25}M_{\odot}$) 
of galaxies from the Padova Millennium Galaxy and Group Catalog (PM2GC) \citep{rosa} and 
the WINGS survey. This work also shows that 
cluster, group and field galaxies at low-$z$ have comparable
mass functions.\footnote{In agreement with the results of \S\ref{mf_envi},
the cut-off masses of low-$z$ mass functions also seem to be different in different 
environments.}
As the next natural step, we investigate now whether the {\it evolution}
of the galaxy stellar mass function changes with environment.

In \cite{morph}, we compared cluster galaxy stellar mass functions at low and
high-$z$ using WINGS and EDisCS data and found a strong
evolution from $z\sim 0.8$ to $z\sim0$, which we attributed primarily to mass growth due to
star formation in late-type cluster galaxies and infalling galaxies. 
We found that the shape at $M_{\ast}> 10^{11} M_{\odot}$ does not evolve 
but the mass function at high redshift is flat below $M_{\ast}\sim 10^{10.8} M_{\odot}$, while
in the Local universe it flattens out at significantly lower masses. 
The population of
$M_{\ast} = 10^{10.2} - 10^{10.8} M_{\odot}$ galaxies must have grown 
significantly between $z=0.8$ and $z=0$.

\cite{pozzetti10}, using data from the zCOSMOS-bright 10k spectroscopic sample, 
quantified the evolution of the mass function in the field since $z\sim 1$. 
They used data from  \cite{baldry08}, who selected galaxies from 
the NewYork University Value-Added Galaxy Catalog
sample as reference at $z=0$. They found a continuous increase with time in the mass function 
for $\log M/M_\ast <11 $ and a much slower increase
at higher masses. 

\begin{figure}[!t]
\centering
\includegraphics[scale=0.4]{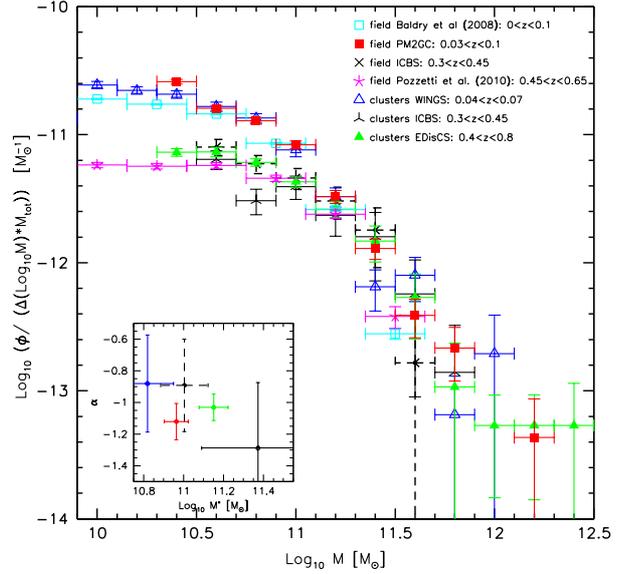}
\caption{Comparison between the mass function in clusters and in
the field at different redshifts.  At low-$z$, red filled squares: PM2GC
(general field); cyan empty squares:   \cite{baldry08}
(general field); blue empty triangles: WINGS (clusters). At intermediate-$z$, black
crosses: ICBS (field); black skeleton triangles: ICBS (clusters). At higher-$z$:
magenta skeleton stars: \cite{pozzetti10} (field); green filled triangles EDisCS (clusters).  
Mass functions are normalised  using the total integrated stellar mass in the range $11.2<\log M_{\ast}/M_{\odot}<11.7$ (the last bin in the ICBS). 
Errorbars  and in-set as in \fig\ref{mf_env2}. 
The evolution of the  shape of the mass function seems not to depend on environment,  
being similar in clusters and in the field.  
However, in all cases the amount of growth as a function 
of cosmic time is different at the low- and high-mass ends.
\label{evol}}
\end{figure}

We are now in the position to compare the evolution of the mass function
in clusters with that in the field.
In \fig\ref{evol}, we examine the evolution from $z\sim 0.4-0.8$ to $z\sim 0$. 
At $z\sim0.07$ we show the mass functions for WINGS clusters and  \cite{baldry08}  and
PM2GC for the general field. At $z\sim0.4$, those of clusters and field from ICBS. 
At $z\sim0.6$, the mass functions shown are those of EDisCS clusters 
and \cite{pozzetti10}  field  (private communication).

Perhaps unexpectedly, the evolution does not depend on global 
environment: it seems similar in clusters and the field.
At similar redshifts, the shape of the mass functions of field and cluster galaxies 
overlap quite significantly, within the errors. 
As cosmic time increases, the number of galaxies at
low and intermediate masses grows with respect to the number of 
massive galaxies. This growth is similar in clusters and in the field.

However, it is hard to quantify accurately the evolution from $z\sim0.4$ and $z\sim0.6$ to $z=0$, 
due to the uncertainties. 
In \tab\ref{tab:fit_global}, best-fit Schechter function parameters are given for 
the WINGS, PM2GC, ICBS and EDisCS
samples. For these, stellar mass estimates and mass functions are derived 
in the same way and the results are directly comparable. 
The best-fit parameters for WINGS and PM2GC agree well within the errors.
The EDisCS and ICBS parameters also agree within the errors, but 
the large uncertainties inherent to the ICBS sample do not allow good constraints of the mass function.
From lower to higher $z$, the characteristic mass $M_\ast^*$ increases, while a clear trend 
for the low mass end slope $\alpha$ is not found.

A direct quantitative comparison with the findings 
of \cite{pozzetti10} and \cite{baldry08} cannot be made since these results 
were obtained using heterogenous data and slightly different redshift
ranges. We stress that comparing stellar mass functions derived 
from independent works can be dangerous, since different 
methods to derive stellar masses which make different assumptions 
can result in differences in the  mass estimates by a factor of $\sim2$--$3$.
These systematic differences can be significant at the high-mass end, 
thus biasing the derived evolution.

\subsection{The mass function in different cluster regions }  
\label{mf_cluster}

\begin{figure*}
\centering
\includegraphics[scale=0.4]{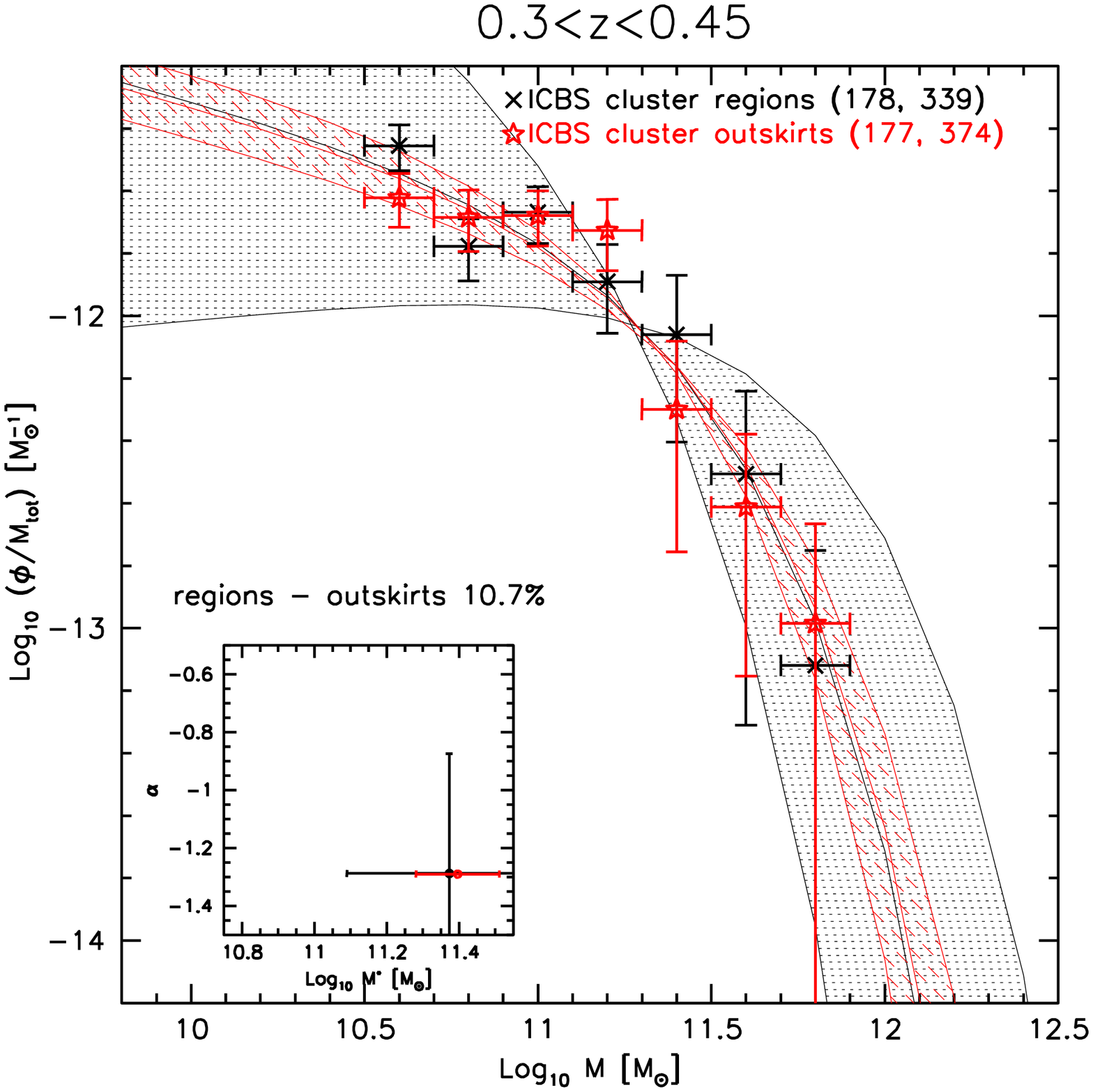}
\includegraphics[scale=0.4]{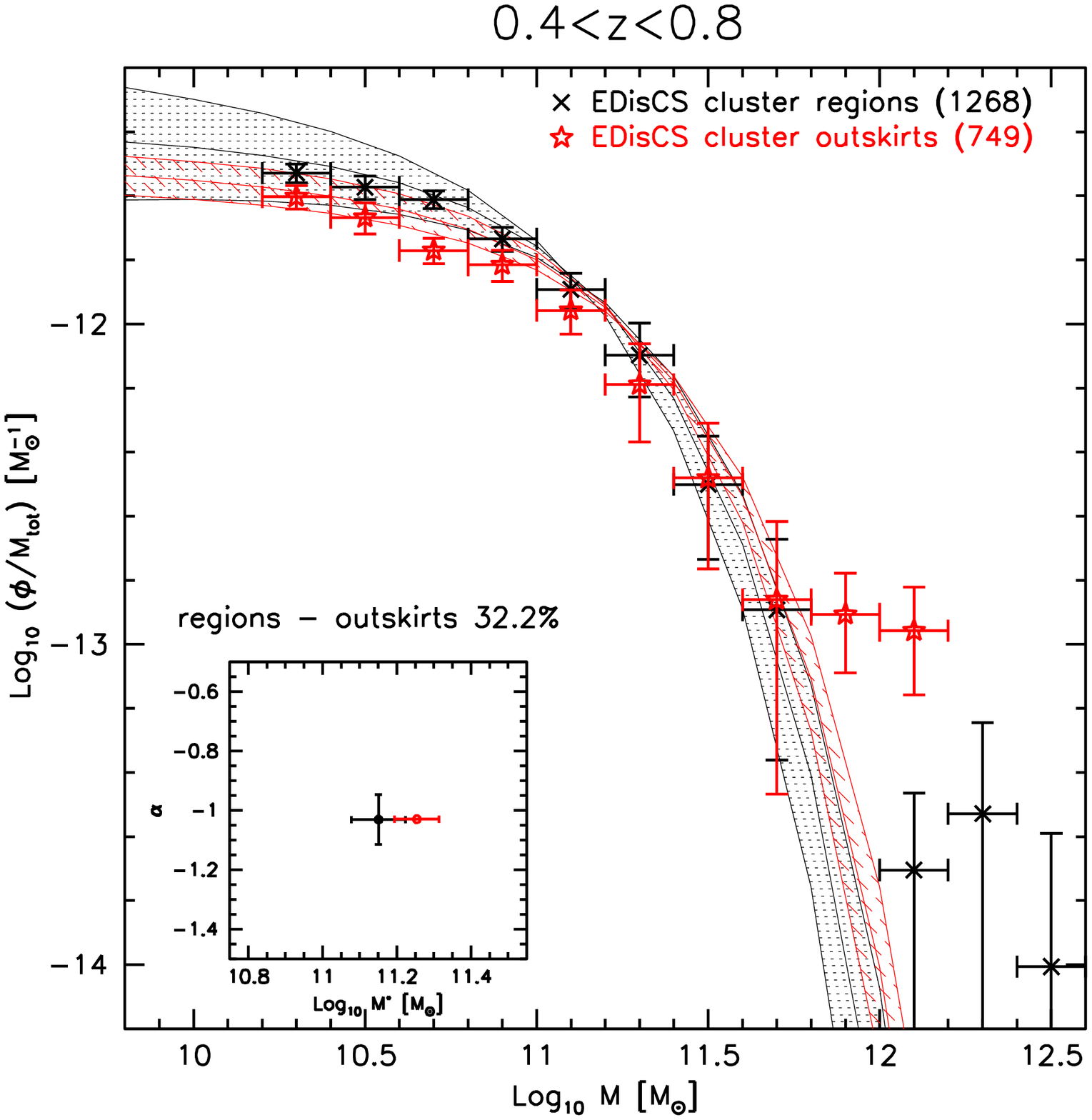}
\caption{Observed mass functions and corresponding Schechter 
function fits for  galaxies  at different clustercentric distances
($R/R_{200} \leq 1$ and $R/R_{200} > 1$) for the ICBS survey (left panel) and
EDisCS (right panel). Black crosses and solid lines represent  cluster regions, and red empty stars and dotted lines the
cluster outskirts.  Mass function normalisations,  errorbars, labels  and  in-sets are as in \fig\ref{mf_env2}.
In both panels, no statistical differences are detected between 
the mass functions of galaxies located at different clustercentric distances. 
\label{mf_RR}}
%\vspace{1cm}
\end{figure*}

We now shift our attention to comparing different environments within clusters. 
As described in \S\ref{def_env}, we subdivide clusters into
regions with different galaxy clustercentric distances.
We first compare in \fig~\ref{mf_RR} the mass functions in 
cluster regions ($R/R_{200} \leq 1$) and the outskirts  
($R/R_{200} > 1$).   
Using both ICBS and EDisCS data, K--S tests are 
unable to detect clearly significant differences 
($P_{\rm K--S}\geq 10\%$ in both cases).
This is also supported by the Schechter function fit results (see \tab\ref{tab:fit_global}).
For this comparison we have fixed the low-mass-end slope $\alpha$ for the outskirts to be 
the same as the one obtained for the cluster regions since it is not well constrained. 
As before, these  fits do not describe adequately the bump observed 
in the EDisCS mass functions at  $M_{\ast} \sim 10^{12.3} M_{\odot}$, 
but they work well at lower masses. 
The fitted parameters indicate that the shape of the mass function is similar
in both regions. This is clearly seen for EDisCS, where the statistics are better, 
but it also seems to be true for the ICBS sample despite the larger statistical uncertainties
(\fig~\ref{mf_RR}).

\begin{figure}
\centering
\includegraphics[scale=0.4]{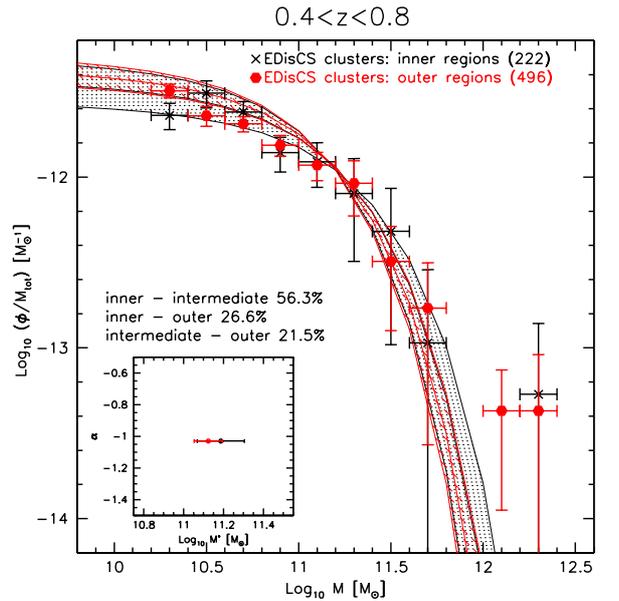}
\caption{Observed mass functions and  Schechter fits for galaxies in different EDisCS cluster regions inside
$R/R_{200}=1$. 
For sake of clarity, only inner  (black
crosses and solid line) and outer
parts (red filled hexagons and dotted line) are plotted. 
Mass function normalisations,  errorbars, labels  and the  in-set  as in \fig\ref{mf_env2}.
The shape of the mass function is similar in the different regions. \label{mf_R}} 
\end{figure}

We next compare three different zones within the cluster virialised regions,
the inner ($0\leq R/R_{200}< 0.2$), intermediate ($0.2\leq
R/R_{200} < 0.6$), and outer ($0.6\leq R/R_{200} <1$) parts.
Both for ICBS data  and EDisCS, the Schechter parameters are compatible within the errors and
the K--S test is always
inconclusive and thus unable to detect any strong variation with
clustercentric distance. To illustrate these results, \fig~\ref{mf_R} shows the mass functions for 
the inner and outer regions of EDisCS clusters. We do not show the plot for ICBS 
data since the quality of the statistics is quite low.

In summary, no overall differences are detected between the cluster virialised regions and their 
outskirts. This agrees with our previous result that the 
global environment does not alter the  shape of the mass distribution since  
the outskirts of clusters can be considered to be 
a transition region between the cluster virialised 
regions and the field, and no differences have been detected between their mass functions. 
Our results agree with those of \cite{vonderlinden10} derived using SDSS data:
excluding BCGs, they found no
evidence for mass segregation in clusters since 
the median mass of cluster galaxies seems to be invariant with cluster radius.

\subsection{The mass function of red and blue galaxies does not depend on environment}\label{mf_colour}

In previous subsections, we have shown that there appears to be no dependence  of the shape
of the mass function on the global environment. 
Our finding is quite surprising because it is known that galaxies located in different
environments have different morphological and star formation
distributions. The question that arises is whether different galaxy
types also follow the same mass distribution.

We subdivide galaxies by colour to separate galaxies with different star formation properties.
Following \cite{peng10}, and after converting to our adopted IMF and the Vega system,
galaxies are assigned to the red sequence if their rest-frame colours obey 
$$
(U-B)_{\rm Vega} \geq 1.10+0.075\log(\frac{M/1.12}{10^{10}M_{\odot}}) -0.18z - 0.88.
$$
The rest of the galaxies are assigned to the blue cloud.

\begin{figure}
\centering
\includegraphics[scale=0.3]{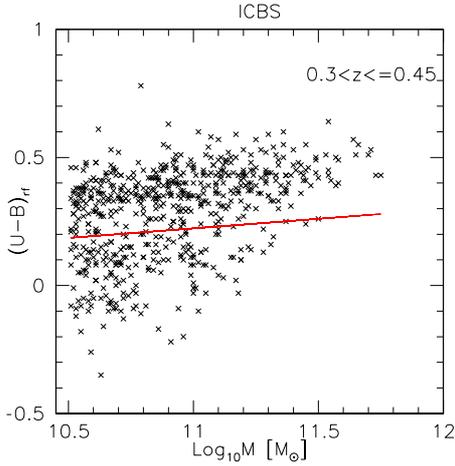}
\caption{Rest-frame $U-B$ colour vs. stellar mass for 
the complete ICBS sample. The line separating red and blue galaxies is
shown. \label{UB}}
\end{figure}

\begin{figure}
\centering \includegraphics[scale=0.4]{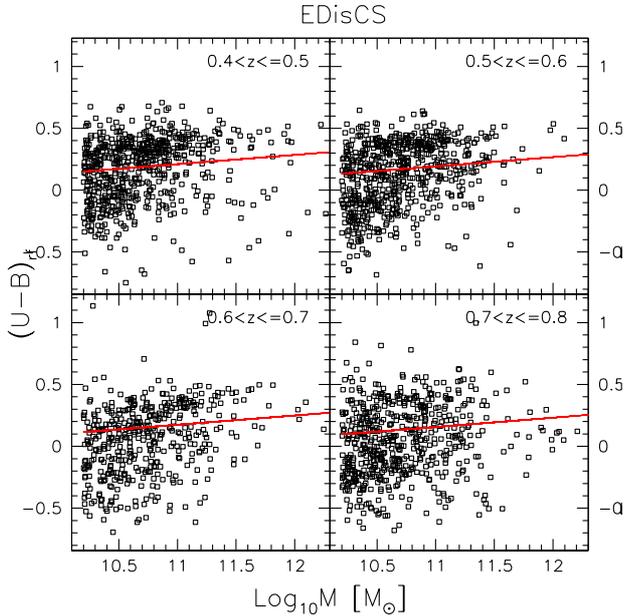}
\caption{
Rest-frame $U-B$ colour vs. stellar mass for the complete 
EDisCS sample split into four redshift bins. The line
separating red and blue galaxies is also shown.\label{UB2}}
\end{figure}

Figures~\ref{UB} and~\ref{UB2} present rest-frame $(U-B)$ colours  as a
function of stellar mass for the ICBS and EDisCS galaxy samples respectively.
The cut adopted to separate
the red and blue populations is shown. Given the broad redshift range spanned by 
EDisCS galaxies we divide this sample into four redshift slices. In doing so 
we avoid the blurring effect of mixing 
galaxies with very different redshifts and the red sequence
and the blue cloud are more easily visible. 

At $0.4<z<0.8$, contamination from dusty star-forming galaxies can be
quite important when selecting red galaxies using only one colour
\citep{wolf09}. Hence, the stellar mass function of red galaxies
is not the stellar mass function of quiescent galaxies only. 
In EDisCS, the level of dusty
star-forming galaxy contamination is $\sim10\%$ for 
our adopted red galaxy selection (estimated following \citealt{brammer11}).
Unfortunately, we cannot make a similar estimate for ICBS given
the limited wavelength coverage of our data.

Tables \ref{frac_icbs} and \ref{frac_ediscs} show the fraction of red
and blue galaxies in both samples. As expected,
these fractions depend strongly on environment.  In the  cluster regions,
red galaxies dominate, especially at high
masses. For $M_{\ast} \geq 10^{10.5} M_{\odot}$, 
$\sim90\%$ of all galaxies in ICBS are red, and $\sim 60\%$  in EDisCS. 
This difference shows the expected redshift evolution,   
If we use the smaller EDisCS mass cut
($M_{\ast} \geq 10^{10.2} M_{\odot}$), we find that for this survey
the red fraction is slightly lower ($\sim 54\%$),
indicating that blue galaxies have preferentially low masses, as 
expected.  In the ICBS sample, the red fraction 
reaches a minimum of 41.7\% in the cluster outskirts.
\begin{table*}[!t]
\centering
\caption{Fractions of blue and red galaxies in the ICBS sample ($0.3<z<0.45$). Errors
are computed as binomial errors. Both observed and
completeness-weighted numbers are listed.\label{frac_icbs}}
\begin{tabular}{lccccc}
\hline
&\multicolumn{4}{c}{ICBS}\\
& \multicolumn{4}{c}{$M_{\ast} \geq 10^{10.5} M_{\odot}$} \\
& \multicolumn{2}{c}{red} 	& \multicolumn{2}{c}{blue} \\ 
\hline
            		&\%$_{\rm obs}$         	& \%$_{\rm w}$	     	&\%$_{\rm obs}$         	& \%$_{\rm w}$\\
\hline	
cluster regions	&91.2$\pm$2.5\%	&92.9$\pm$1.4\%  		&8.8$\pm$2.5\%	&7.1$\pm$1.4\%	\\
cluster outskirts	&41.7$\pm$3.2\%	&40.4$\pm$2.1\%  		&58.3$\pm$3.2\%	&59.6$\pm$2.1\%	\\
groups		&67.7$\pm$5.3\%    &67.3$\pm$3.5\%  		&32.3$\pm$5.3\%	&32.7$\pm$3.5\%	\\
field 		&58.3$\pm$3.3\%	&57.8.$\pm$2.1\%		&41.7$\pm$3.3\%	&42.2$\pm$2.1\%	\\
\hline
\end{tabular}
\end{table*}

\begin{table*}
\centering
\caption{Fractions of blue and red galaxies in the EDisCS
sample ($0.4<z<0.8$). Errors are computed as binomial errors.  Results 
using both the EDisCS
intrinsic mass limit ($M_{\ast} \geq 10^{10.2} M_{\odot}$) and the ICBS one
($M_{\ast} \geq 10^{10.5} M_{\odot}$) are
shown.\label{frac_ediscs}}
\begin{tabular}{lccccc}
\hline
&\multicolumn{4}{c}{EDisCS}\\
& \multicolumn{2}{c}{$M_{\ast} \geq 10^{10.2} M_{\odot}$} &  \multicolumn{2}{c}{$M_{\ast} \geq 10^{10.5} M_{\odot}$} \\
& red &blue & red 	& blue \\ 
\hline
            		&\%        	& \%	     	&\%         	& \%\\
\hline	
cluster regions	&54.1$\pm$1.4\%	&45.9$\pm$1.4\%  		&60.0$\pm$2.0\%		&40.0$\pm$2.0\%	\\
cluster outskirts	&36.8$\pm$1.8\%	&63.1$\pm$1.8\%		&41.5$\pm$2.3\%		&58.5$\pm$2.3\%	\\
groups		&40.3$\pm$1.8\%    &59.7$\pm$1.8\%  		&43.0$\pm$2.6\%		&57.0$\pm$2.6\%	\\
\hline
\end{tabular}
\end{table*}

The fraction of red galaxies depends strongly on mass in both samples. 
Red galaxies dominate the high mass end in any environment. 
These fractions (see Table \ref{frac_rosse}) are consistent within 1--2$\sigma$ with the fractions 
of quiescent galaxies  obtained from the number densities given by \cite{brammer11} for field 
galaxies at $z\sim 0.613$: 
$50 \pm 12\%$ for $\log M_\ast/M_\odot = 10.2-10.6$, $75 \pm 19\%$ for $\log M_\ast/M_\odot =10.6-11.0$, and 
$92 \pm 33\%$ for $\log M_\ast/M_\odot =11.0-11.6$.\footnote{The specific 
way in which passive and star-forming galaxies are selected strongly affects their 
measured fractions. Therefore, comparisons between different studies need to be 
taken with care. Note also that \cite{brammer11} study field galaxies in a redshift range 
slightly higher than the ICBS one.}

\begin{table}
\centering
\caption{Red galaxy fractions as a function of mass both in the EDisCS and ICBS samples. 
Errors are computed as binomial errors.  For ICBS, only weighted fractions are given.\label{frac_rosse}}
\begin{tabular}{lccc}
\hline
 $\log M_\ast/M_\odot =$		&$10.2$--$10.6$&$10.6$--$11$	&$11.0$--$11.6$\\
\hline
\multicolumn{4}{c}{{\bf ICBS} }\\
clusters	&							&	$90.1\pm 2.9 \%$	&	$97.7\pm 2.5\%$ \\	
outskirts	&							&	$	59.3\pm 3.9 \%$	&	$	78.1\pm 4.3	\%$ \\
groups	&							&	$	64.6\pm 5.1 \%$	&	$	83.0\pm 6.3\%$ \\
field		&							&	$	51.0\pm 2.9 \%$	&	$	79.7\pm 3.6\%$ \\
\hline
\multicolumn{4}{c}{{\bf EDisCS} }\\
clusters	& $44.7\pm 2.2 \%$	&	$	59.4\pm 2.5 \%$	&	$	67.5\pm 3.3\%$ \\
outskirts	& $28.5\pm 2.5 \%$	&	$	44.4\pm 3.4 \%$	&	$	41.7\pm 4.3\%$ \\
groups	& $34.5\pm 3.0 \%$	&	$	43.8\pm 3.7 \%$	&	$	46.0\pm 4.6\%$ \\
\hline
\end{tabular}
\end{table}

\begin{figure*}
\centering
\includegraphics[scale=0.28]{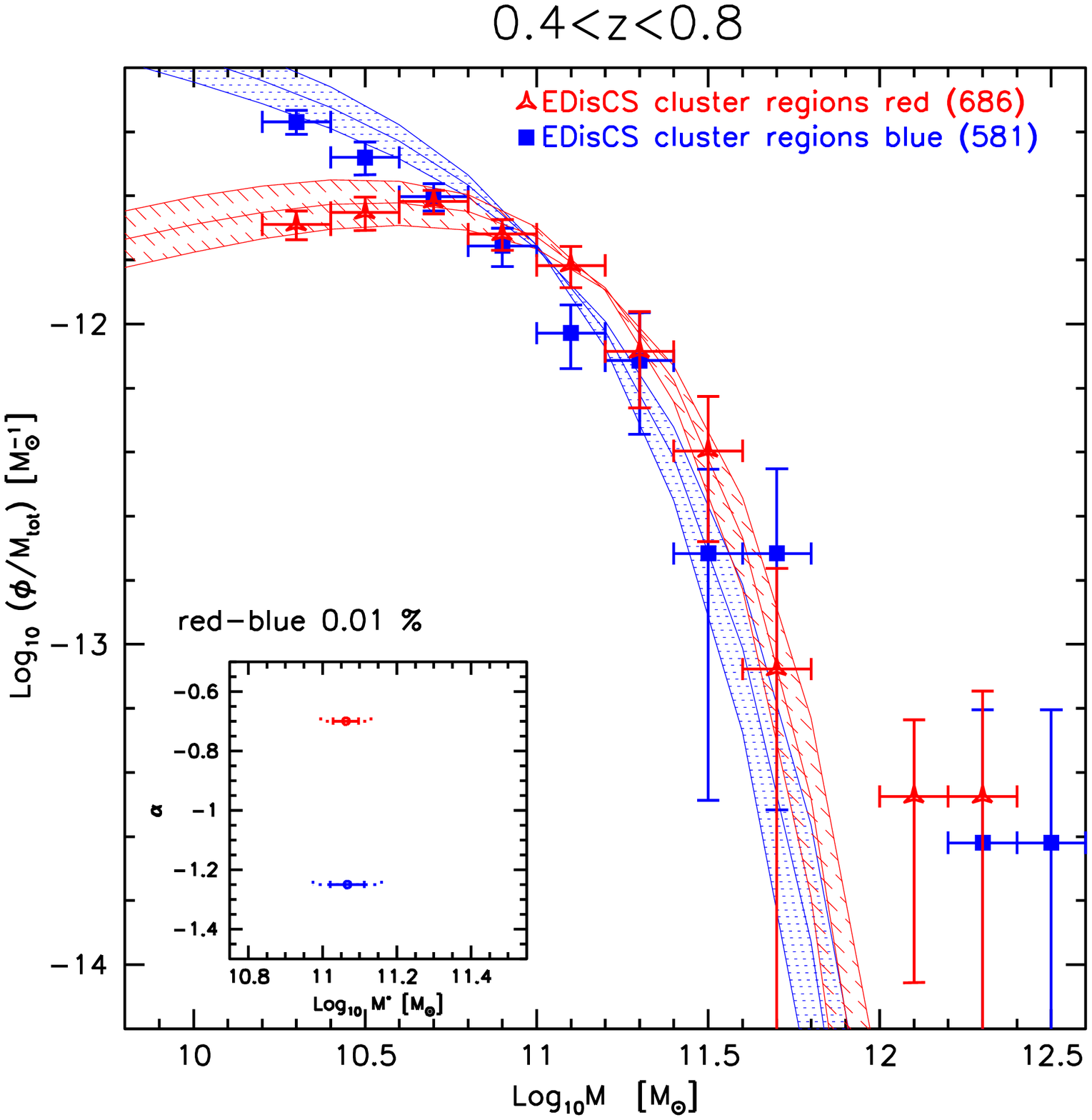}
\includegraphics[scale=0.28]{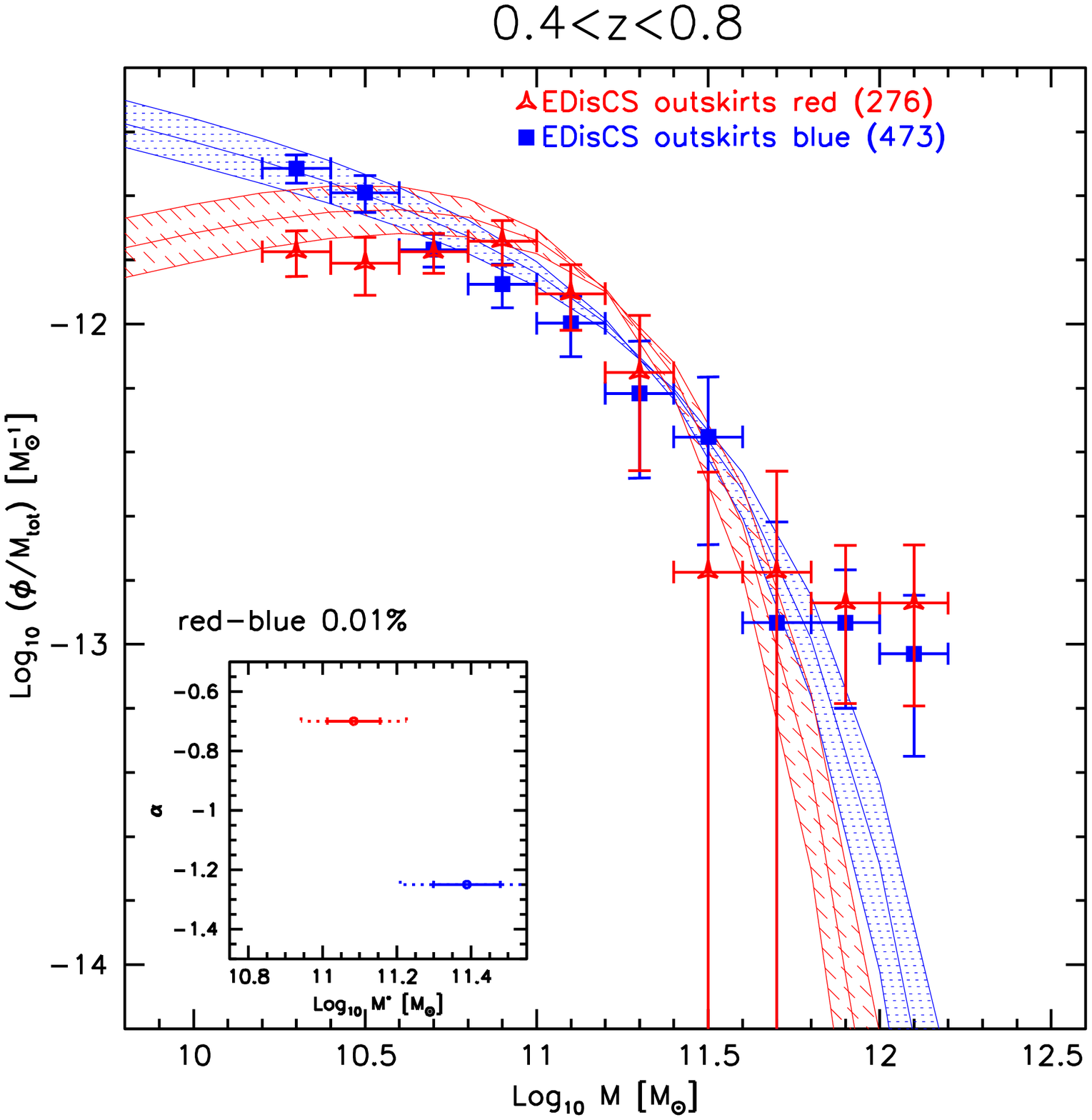}
\includegraphics[scale=0.28]{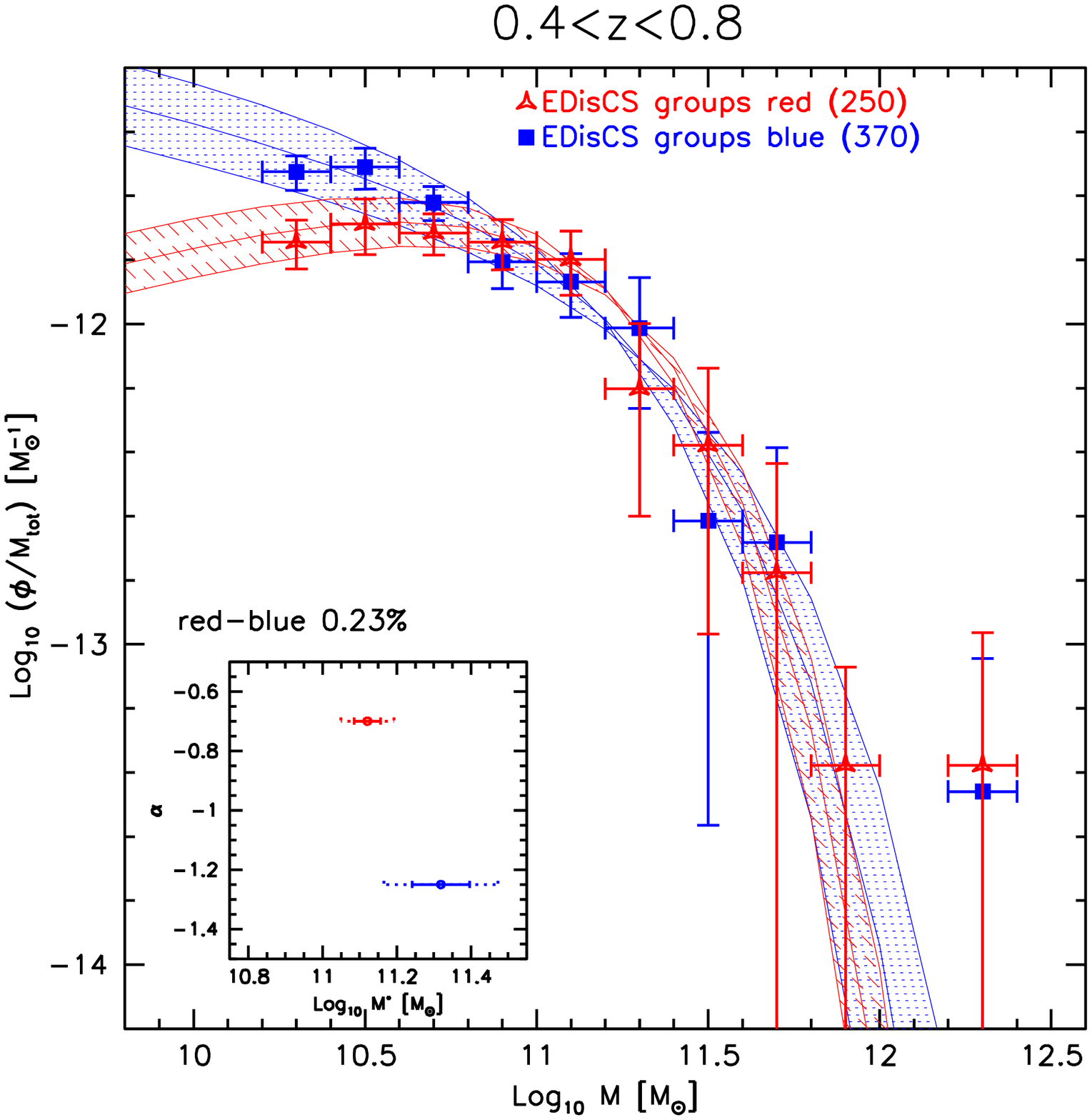}
\caption{Observed mass functions and Schechter fits for blue and red EDisCS galaxies in clusters (left panel), 
in cluster outskirts (central panel) and  in groups (right panel). Blue filled squares and solid lines correspond 
to  blue galaxies, and red empty triangles and dotted lines to red galaxies. 
Mass function normalisation,  errorbars, labels  and  in-sets as in \fig\ref{mf_env2}.
In the in-sets, errorbars represent the $1\sigma$  (solid line) and  $2\sigma$ (dotted line) errors. 
In clusters, in cluster outskirts, and in groups blue and red galaxies have different mass distributions.
\label{rb_ediscs}}
\end{figure*}

In \fig~\ref{rb_ediscs} we show the mass function of red and blue galaxies using EDisCS data. 
For $M_{\ast} \geq 10^{10.2} M_{\odot}$, red and blue galaxies have 
different mass distributions in all the environments considered (clusters, cluster outskirts and groups),
Visual inspection and K--S tests indicate these differences are highly significant in all cases.
The blue galaxy population tends to contain proportionally more low mass galaxies than the red one. 
This is particularly evident in  clusters (both virialised regions and outskirts). 
The blue mass function is thus steeper than the red one. 

ICBS red and blue galaxies also have clearly different mass functions in the field and outskirts (plots not shown).
However, the poor statistics of the ICBS cluster sample prevent the detection of any differences
even if present at the same level as in EDisCS clusters. We tested this by using the EDisCS 
cluster sample to build 1000 Monte-Carlo samples of cluster galaxies with the same numbers and selection
criteria as ICBS. In the majority of cases, K--S tests fail to detect significant 
differences between the mass functions of red and blue galaxies, even though they do exist.  
Moreover, adopting the higher ICBS mass limit ($M_{\ast} \geq 10^{10.2} M_{\odot}$) 
for the EDisCS cluster galaxies we still find
that red and blue galaxies follow
different mass distributions ($P_{\rm K--S}=0.18$\%). Therefore,  
the differences are not limited to those found at lower masses.

\begin{figure*}
\centering
\includegraphics[scale=0.4]{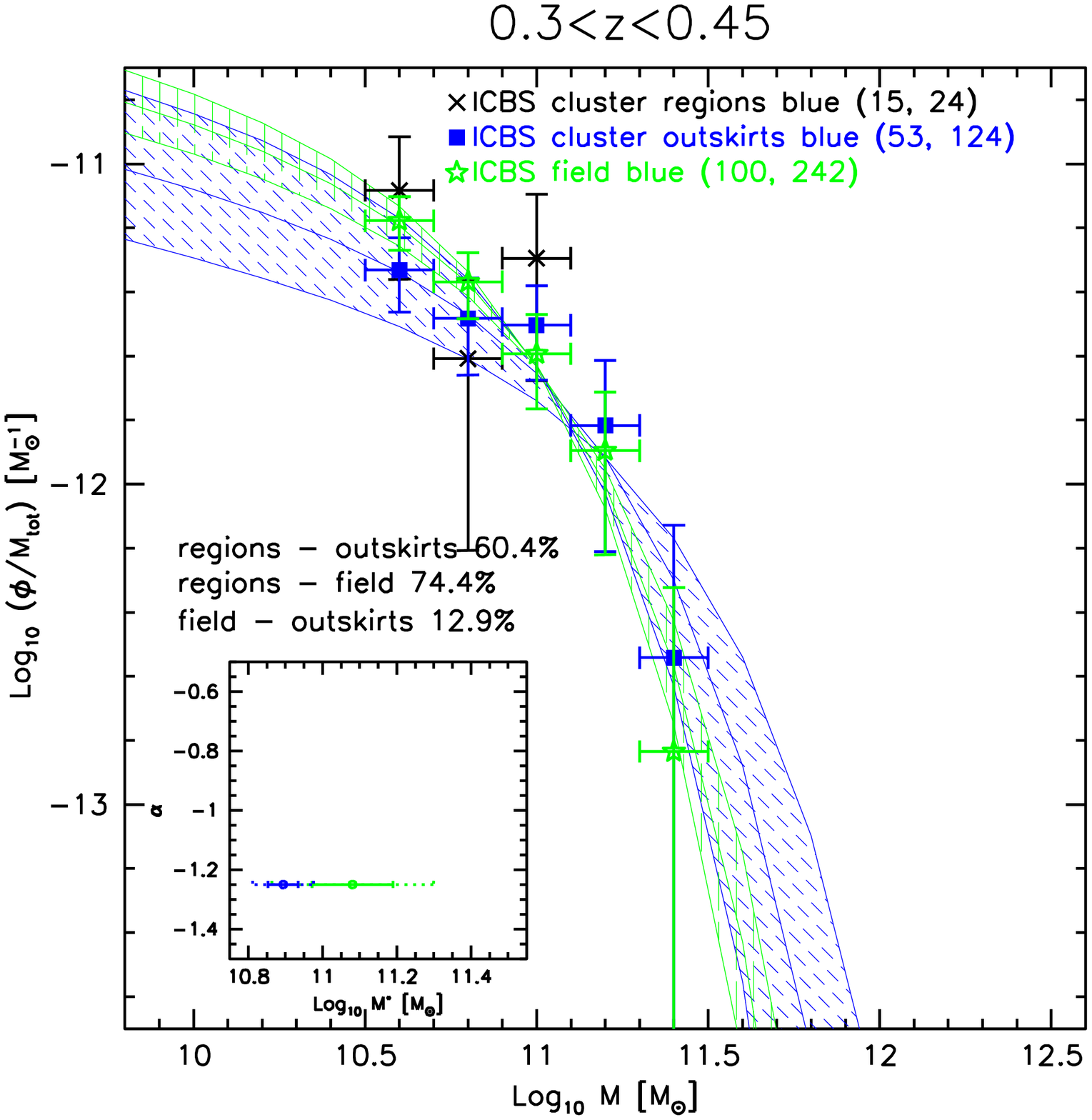}
\includegraphics[scale=0.4]{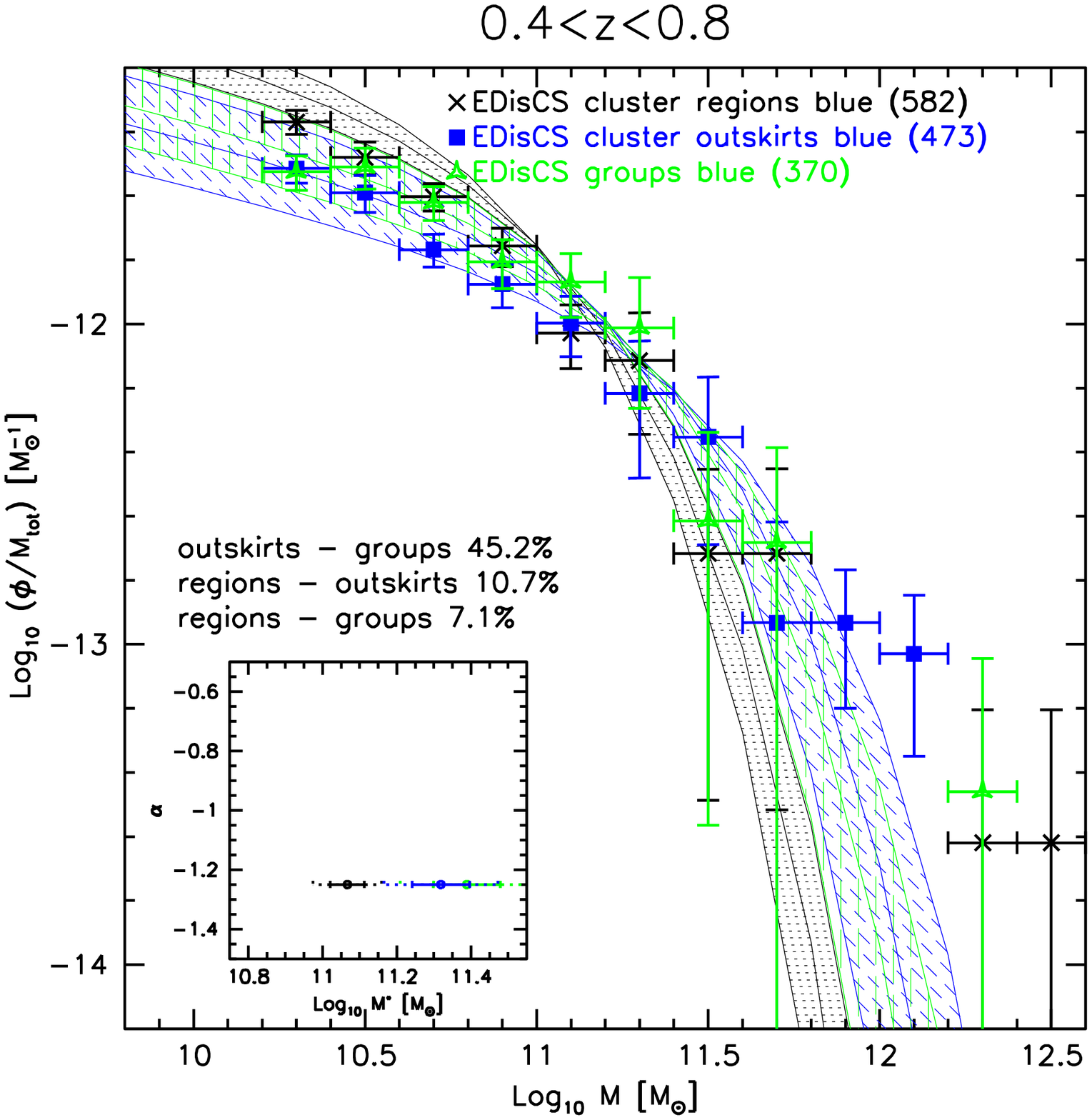}
\caption{Observed mass function and Schechter function fits for blue galaxies in different environments
for the  ICBS (left panel) and EDisCS (right panel) samples. Black crosses and solid lines represent
the cluster regions, blue filled squares and dotted lines the cluster outskirts, and green empty
triangles and dashed lines the groups. The mass distribution of ICBS blue cluster galaxies 
could not be fitted given the small number of bins.
Mass function normalisations,  errorbars, labels  and  in-sets as in \fig\ref{mf_env2}.
In the in-sets, errorbars represent the $1\sigma$ (solid line) and  $2\sigma$ (dotted line) errors. 
For blue galaxies, no differences can be detected between the 
mass functions of galaxies located in clusters, groups, and in the field.
\label{blue}}
\end{figure*}

\begin{figure*}
\centering
\includegraphics[scale=0.4]{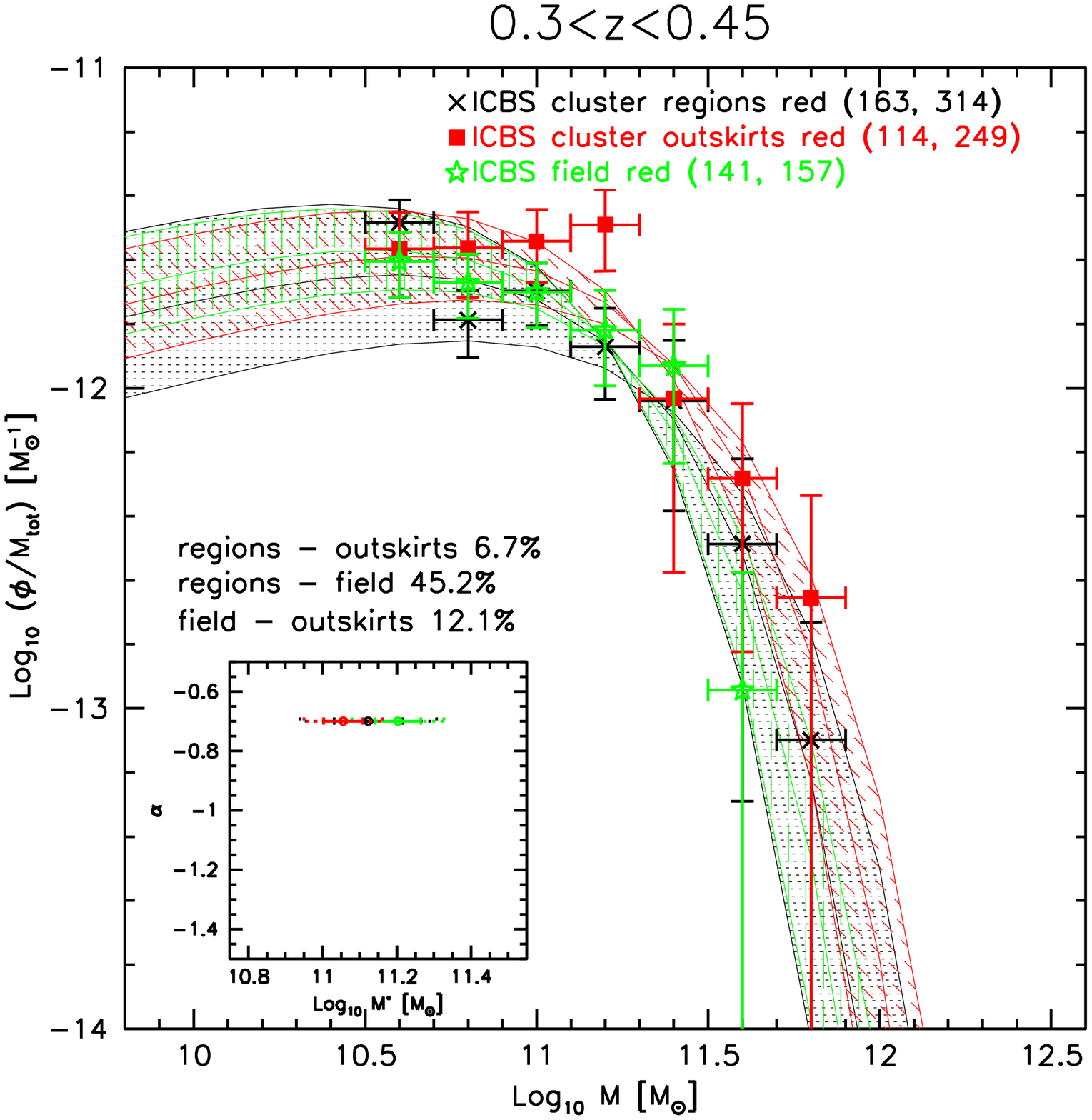}
\includegraphics[scale=0.4]{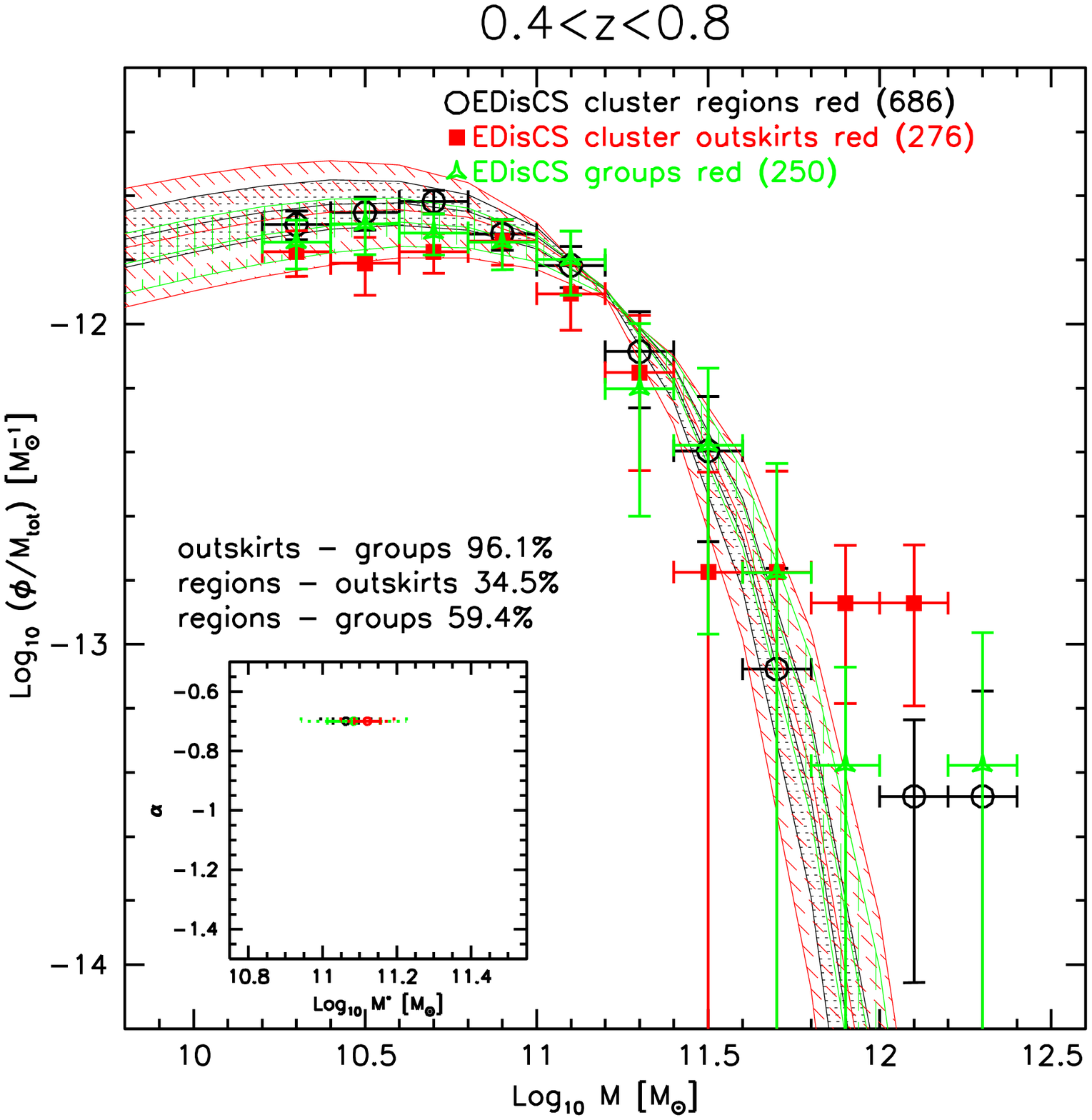}
\caption{
Observed mass function and Schechter function fits for red galaxies in different environments
for the  ICBS (left panel) and EDisCS (right panel) samples. Black crosses and solid lines represent
the cluster regions, red filled squares and dotted lines the cluster outskirts, and green empty
triangles and dashed lines the groups. 
Mass function normalisations,  errorbars, labels  and  in-sets as in \fig\ref{mf_env2}.
In the in-sets, errorbars represent the $1\sigma$ (solid line) and  $2\sigma$ (dotted line) errors. 
For red galaxies, no differences can be detected between the 
mass functions of galaxies located in clusters, groups, and in the field.
\label{red}}
\end{figure*}

\begin{table*}[!t]
\caption{Best-fit Schechter function parameters ($M_\ast^{*}$, $\alpha$,  $\phi^{*}$) for the mass functions of 
galaxies in different environments and with different colours.\label{tab:fit_col}}
\begin{center}
%\centering
\begin{tabular}{llccr}
\hline
	&	&	$\log M_\ast^*/M_\odot$ & 	$\alpha$ &$ \log \Phi^*$\\
\hline	
{\bf ICBS} &	cluster regions red	&$11.12\pm0.09$ & 	$-0.70$ 	&	 $62\pm9$ \\
	&	cluster regions blue	&--- & 	---	&	 --- \\
	&	cluster outskirts red &$11.20\pm0.06$&	$-0.70$	&	$47\pm5$ \\
	&	cluster outskirts blue & $11.08\pm0.11$&	$-1.25$	&	 $22\pm5$ \\
	&	field red		& $11.06\pm0.05$	&	$-0.70$	&      	 $75\pm7$\\
	&	field blue		& $10.89\pm0.04$ 	&	$-1.25$	&     $66\pm7$\\
\hline	
{\bf EDisCS} &   	cluster regions red	&$11.06\pm0.03$ &	$-0.70$ 	&	 $114\pm4$ 	\\
	& cluster regions blue & $11.07\pm0.05$ &	$-1.25$	&	 $57\pm4$ 	\\
	&	cluster outskirts  red   &$11.20\pm0.09$	 &	$-0.70$ 	&	 $47\pm5$  \\
	&	cluster outskirts  blue   &  $11.39\pm0.09$ &	$-1.25$	&	 $29\pm3$ \\
	&	groups red 	& $11.12\pm0.04$ &	$-0.70$	&	  $39\pm1$  \\
	&	groups blue	& $11.32\pm0.08$  &	$-1.25$ &	 $25\pm3$   \\
\hline
\end{tabular}
\end{center}
\end{table*}

Finally, we compare separately the mass function of blue
(\fig~\ref{blue}) and red  (\fig~\ref{red}) galaxies in different
environments.
Visual inspection of the plots and K--S tests provide no indication that 
the environment affects the mass function of blue and red galaxies.  
We also derive best-fit Schechter function parameters (\tab\ref{tab:fit_col}) to the observed mass functions. 
Since the data lack the necessary number statistics to allow robust estimation of the faint end 
slope $\alpha$, we assume fixed values for the red and blue samples separately. 
For the red galaxies we use  $\alpha  = -0.7$  (the same value found by \citealt{borch06}),  
For the blue galaxies we use $\alpha  = -1.25$, The value used by \cite{borch06} ($\alpha=-1.45$) 
is too high to describe well our mass functions, so  our choice is based on the value obtained 
by free-fitting the EDisCS mass functions. Due to the high mass limit and the low number statistics, 
we are unable to find a good fit to the mass function of the blue galaxies in the ICBS clusters. 
As before, we note that a Schechter fit does not describe well the highest mass end of the EDisCS galaxies.

The analysis of the fit parameters indicates that in all environments red and blue galaxies
have comparable values of $M_\ast^*$ (within $1$--$2\sigma$ errors). 
Our findings are in agreement with  \cite{borch06} and \cite{ilbert10}. These authors 
also found that at intermediate redshifts galaxies of different colour show rather similar $M_\ast^*$. Explicitly, 
at $z\sim0.5$, \cite{borch06} found that red galaxies have $\log M_\ast^*/M_\odot ={10.95\pm0.10}$ and blue galaxies   
${10.93\pm0.12}$. Similarly, \cite{ilbert10} found that red sequence galaxies have $\log M_\ast^*/M_\odot  ={10.97\pm0.03}$ and intermediate activity galaxies  ${10.93\pm0.03}$. In summary,  we find that galaxies of the same colour are described by  mass functions 
with similar shape and  $M_\ast^*$ in all environments. This conclusion is  particularly robust for the EDisCS survey, given its high number statistics.

\section{Discussion}\label{disc}
The results from the EDisCS and ICBS samples 
presented in previous sections are generally in
qualitative agreement (or, at the very least, compatible within the uncertainties).  
Unfortunately, given the different characteristics of both surveys (different redshift ranges and
selection criteria), a direct quantitative comparison is not possible. 
In both surveys, by sampling the mass functions in 
a broad range of global environments, we
are effectively studying galaxies in dark matter haloes with 
a wide range of masses. Despite this, we find no obvious dependence of 
the galaxy mass distribution on global environment. Explicitly,  
galaxies located in clusters, groups, and the field seem to follow mass distributions characterised by similar shapes.
This result is surprising and, at some level, perhaps contrary to most expectations.

It will be useful to compare our observational results
with theoretical expectations
trying to understand whether
simulations predict any mass segregation with environment, considering 
both the initial and evolved halo
masses, and their evolution with redshift 
as a function of environment. This will be the 
subject of a future paper (Vulcani et al. in preparation). 
In this section we will concentrate on the empirical results
trying to clarify the emerging picture.

\subsection{The evolution of the mass function in different environments}\label{ev}
In \cite{morph} we
argued that the evolution observed in clusters is driven by the
mass growth of galaxies caused by star formation in both cluster galaxies
and, more importantly, in galaxies infalling from the cluster surrounding
areas. In that preliminary analysis carried out with inhomogeneous
data, we also hypothesized that infalling galaxies could perhaps follow a
steeper mass distribution  than cluster galaxies. If that were the case, infalling 
galaxies would contribute more to the
intermediate-to-low mass population. However, we found no evidence of
any difference between the cluster and field mass functions taken from the literature. 
In the present work we have analysed the mass functions of galaxies in the field and, 
most importantly, in the cluster surrounding areas in a self-consistent way. We 
have thus been able to characterise the  mass distribution of galaxies 
that are supposed to fall into clusters 
and compare it to the cluster mass function at the same redshift. 
We have found that for the mass ranges considered ($\log M_{\ast}/M_{\odot} \geq 10.5$), the
mass function is invariant with environment: galaxies located in different
environments contain very similar mass distributions. We can thus conclude  
that the observed evolution of the mass function in clusters \citep{morph} can probably  not 
be explained by galaxies of different masses residing in different environments. This seems to
be the case at least for the relatively high galaxy masses studied here. 
For these galaxies, star formation is the only major process 
left to explain the observed mass growth, both in clusters and in the field.

Moreover,  by analysing also the field mass function of galaxies in the local universe 
(Calvi et al. in preparation), we have investigated the evolution of the  field mass function 
from $z\sim 0.4$ to $z\sim 0$ and compared this to the evolution found 
in clusters \citep{morph}. Our results show that 
for $\log M_{\ast}/M_{\odot} \geq 10.2$ the
galaxy stellar mass function evolves in the same way in all environments. In all cases,
it becomes steeper with time. Thus, the number of 
intermediate-mass galaxies grows proportionally at the same rate in clusters and in the field.

Our findings are quite surprising since it is well known that galaxies in different
environments and  with different
stellar masses have different star formation properties and are subject to different physical processes.
We expected that the processes responsible for suppressing or halting star formation
would be different in clusters and in the field, resulting in 
different mass growth rates and timescales in different environments. 
What we find instead is that at the redshifts considered most of the galaxy mass 
appears to have already been assembled, and that environment-dependent
processes have had no significant influence on galaxy mass.

\subsection{The blue and red mass functions}\label{rb}

In \S\ref{mf_colour} we found that in all environments red and
blue galaxies have different mass functions:
blue galaxies have steeper mass functions than red ones
(\fig~\ref{rb_ediscs}). However, we also found that, separately, 
blue and red galaxies have very similar mass function in 
all environments (see \fig~\ref{blue}
and \fig~\ref{red}). Additionally, the fraction of red and blue galaxies
strongly depends on environment: red galaxies
dominate the cluster regions while blue galaxies are found mostly
in the  field. It is therefore quite surprising that the total mass 
function is almost always the same in all environments.  
 
\begin{figure*}
\centering
\includegraphics[scale=0.25]{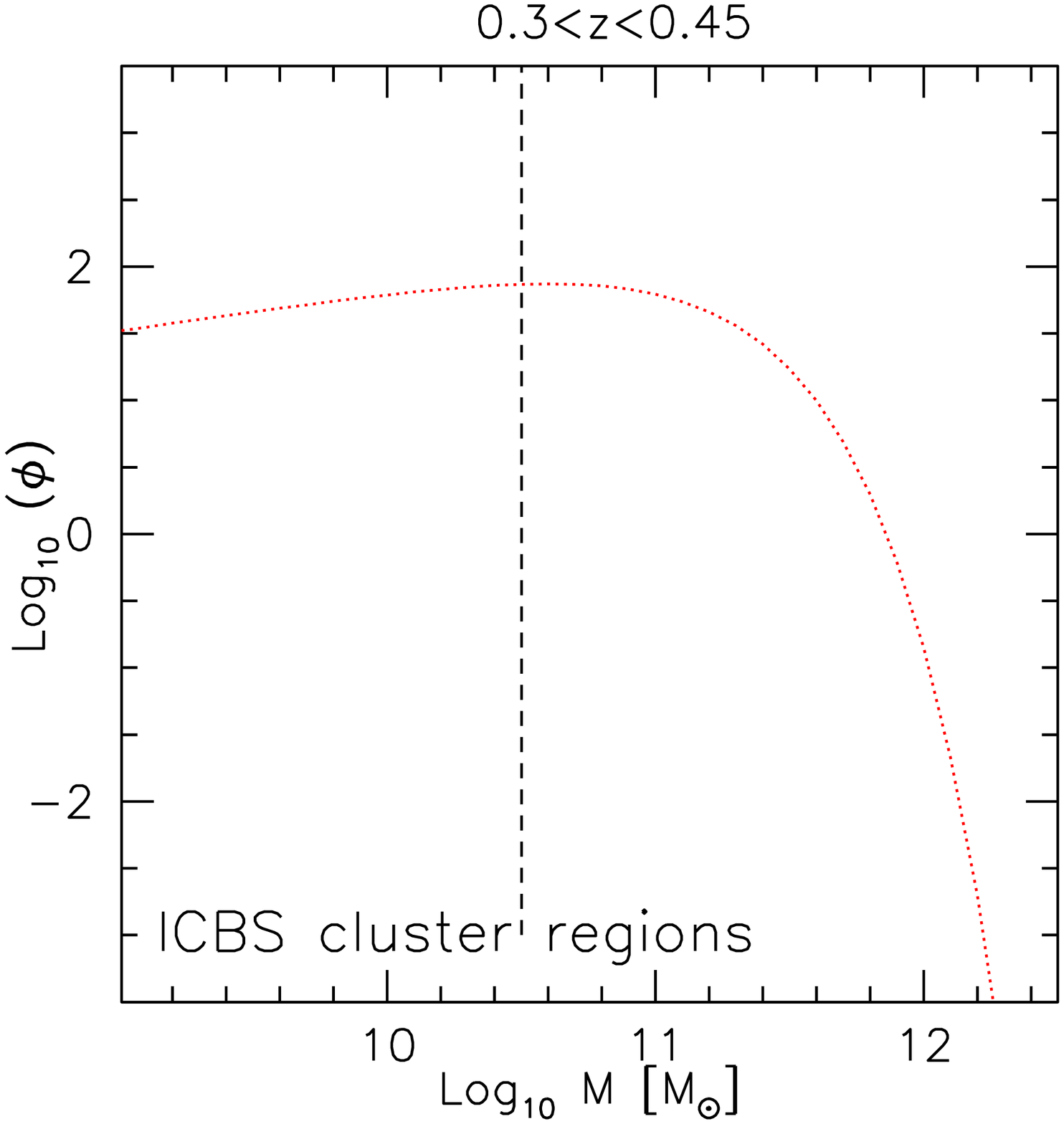}
\includegraphics[scale=0.25]{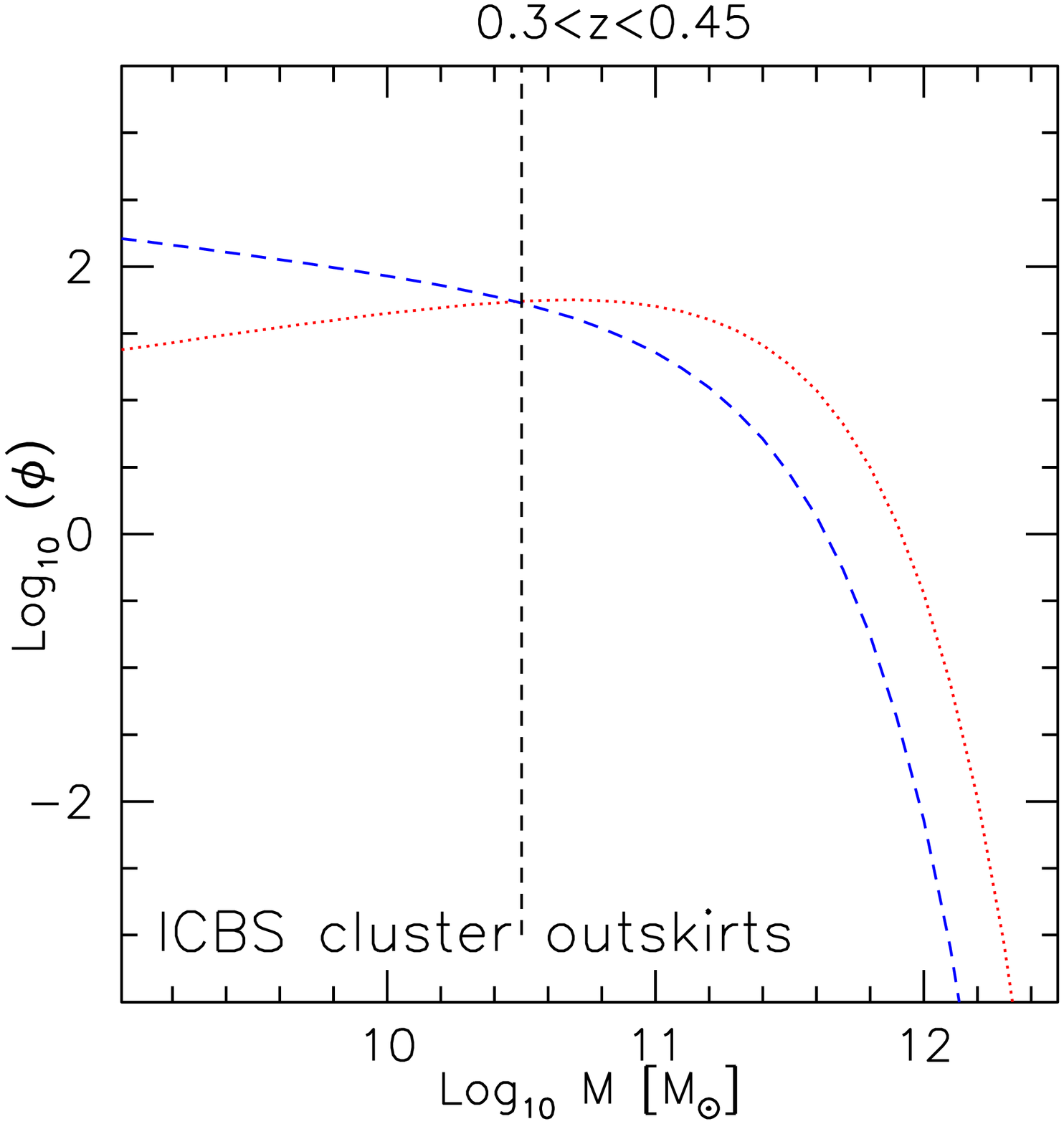}
\includegraphics[scale=0.25]{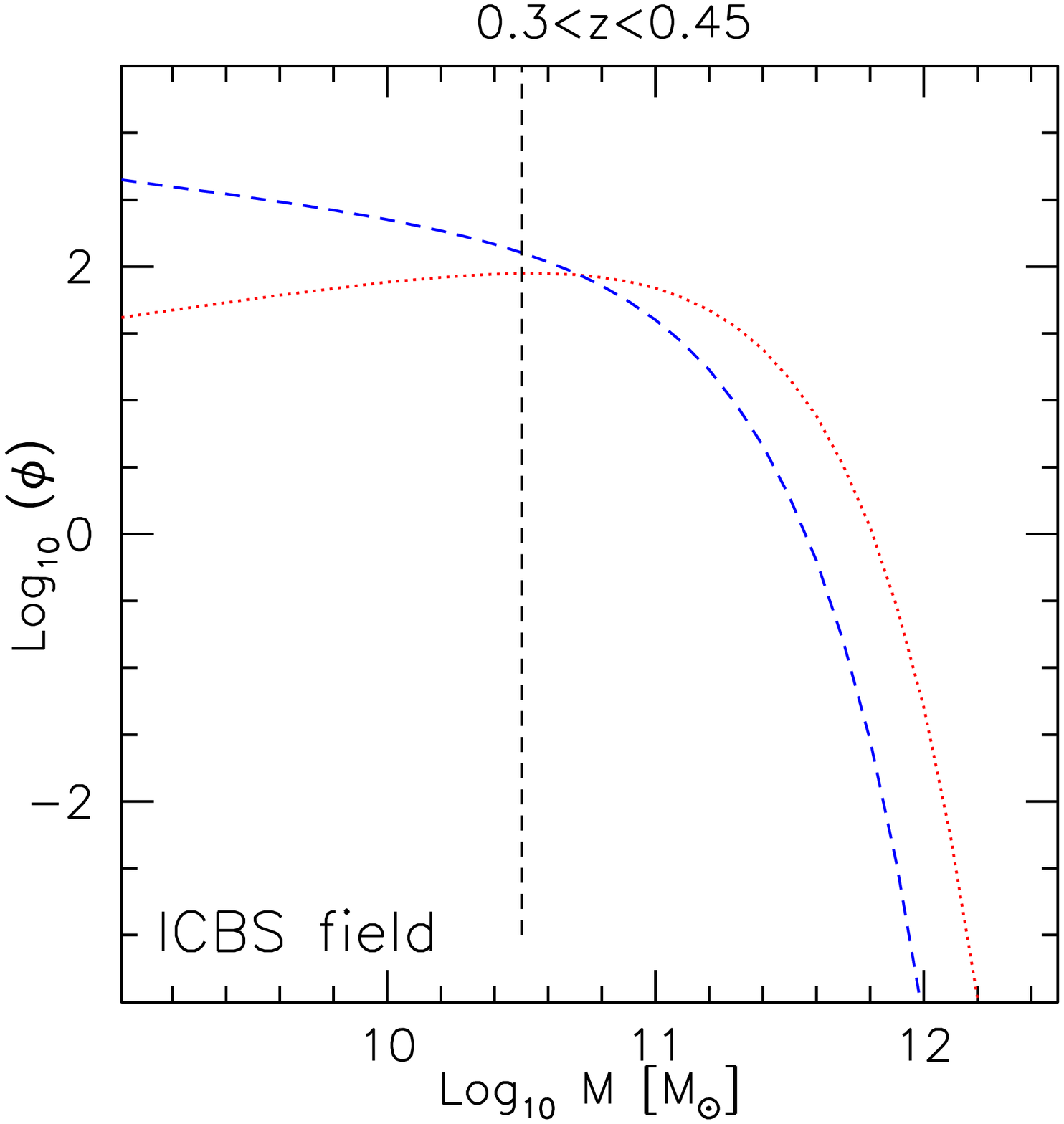}
\includegraphics[scale=0.25]{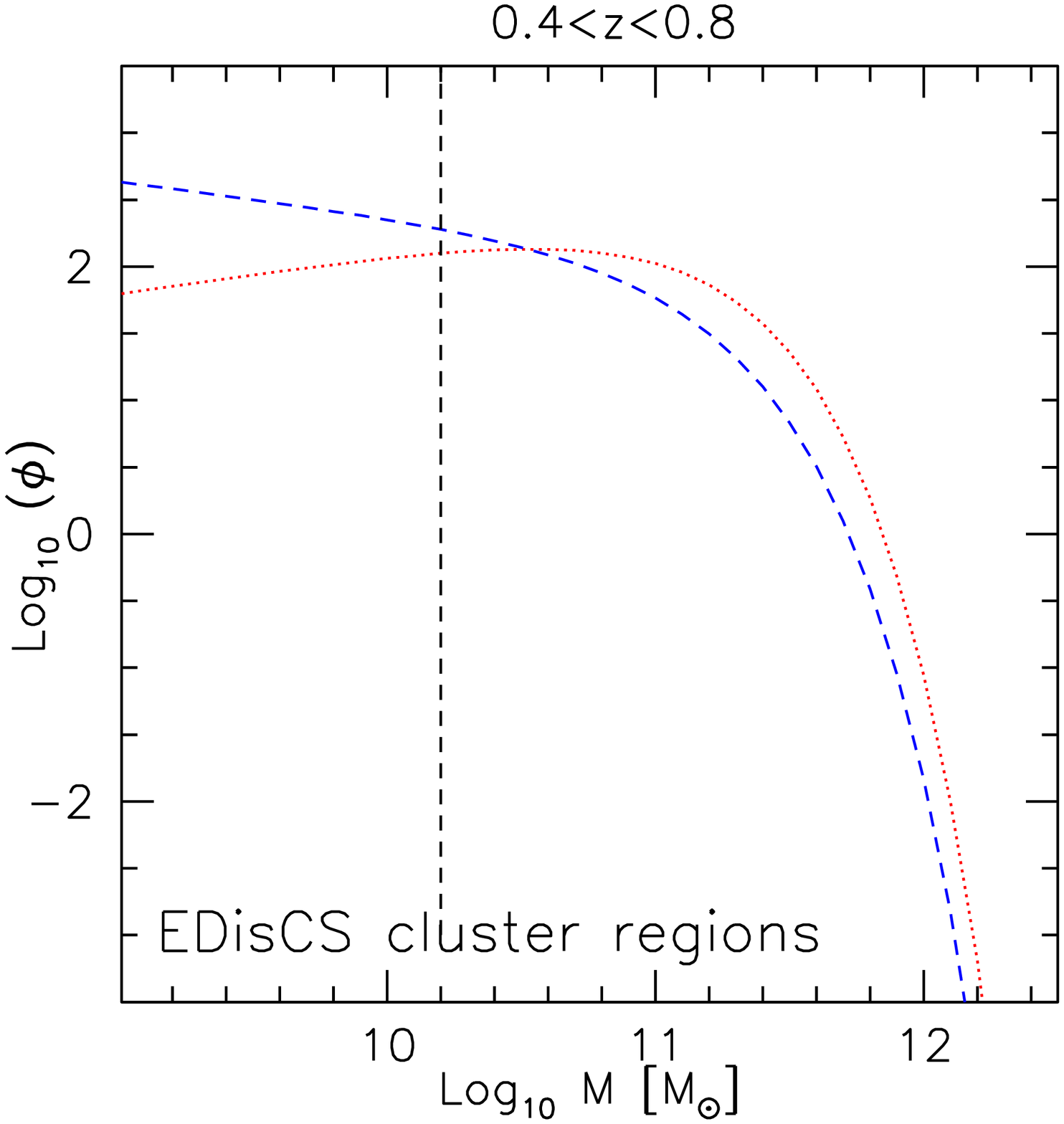}
\includegraphics[scale=0.25]{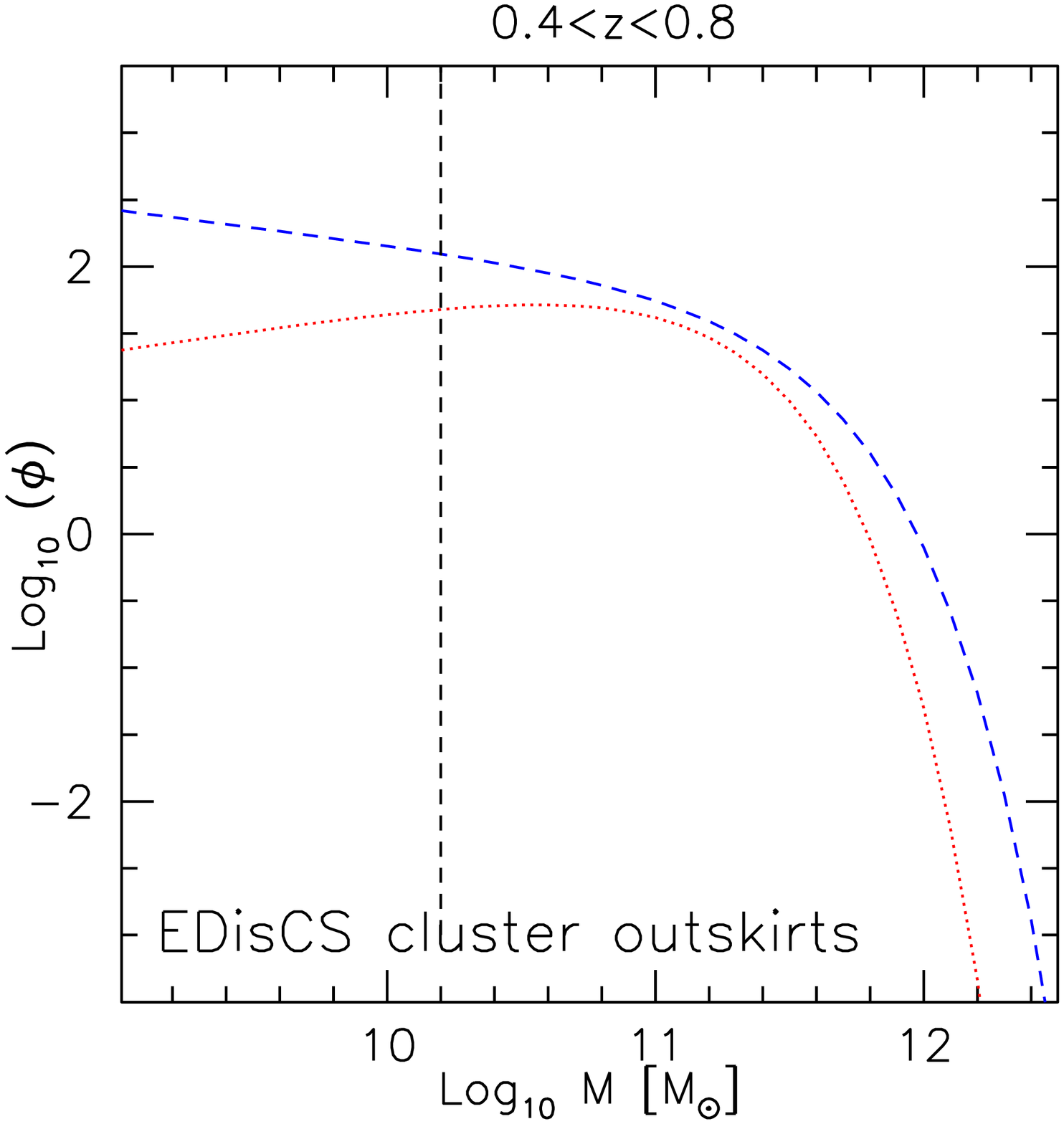}
\includegraphics[scale=0.25]{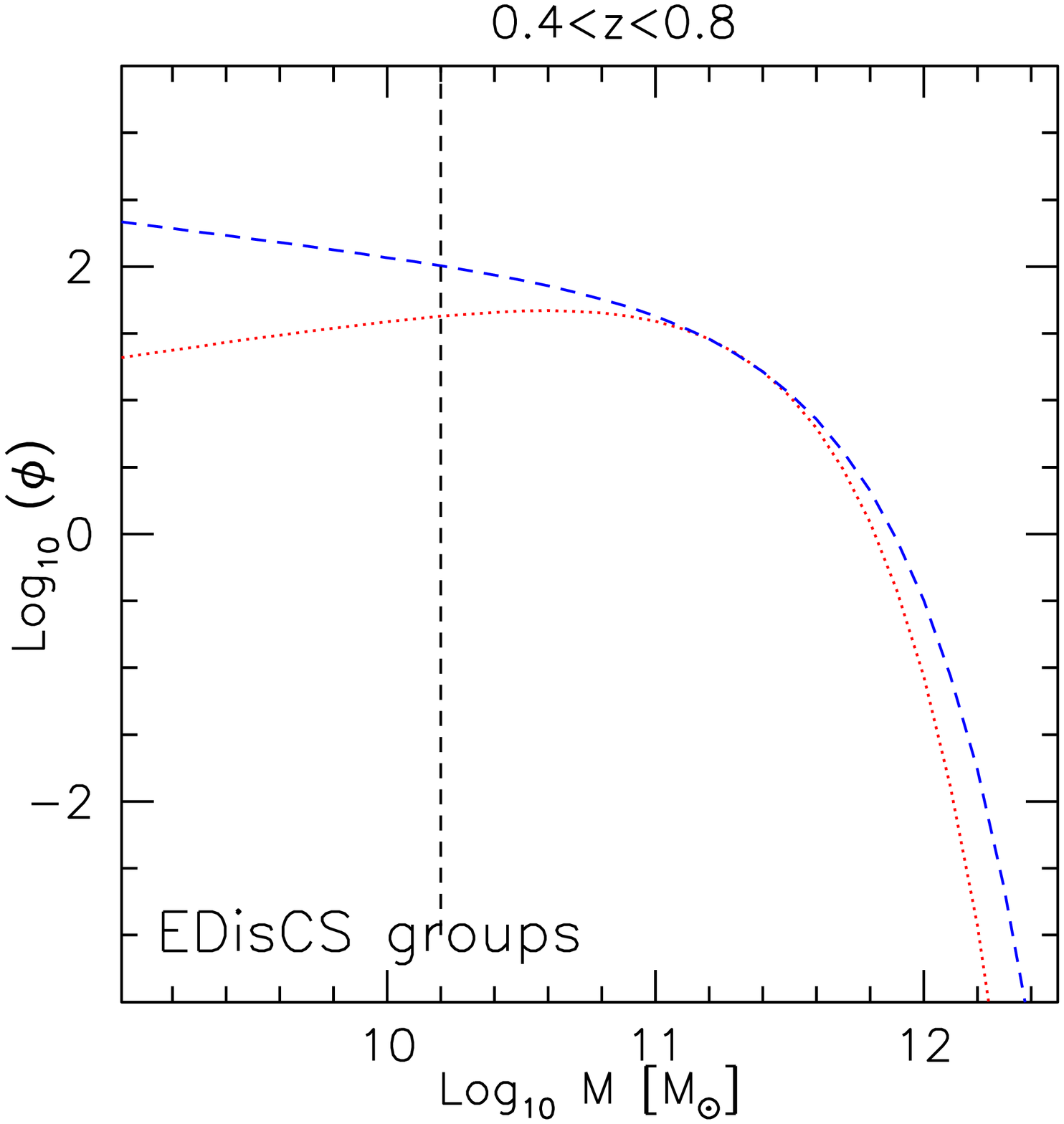}
\caption{Schechter fits for the blue (blue dotted lines) and red (red dashed lines) galaxies in the ICBS  (upper panels) and EDisCS (lower panels) 
samples in clusters (left panels), the outskirts (central panels) and  groups (right panel for EDisCS) and the field (right panel for ICBS). 
Black vertical dashed lines represent the range in stellar mass probed by these surveys. The curves are not normalised.
The blue ICBS cluster mass function is omitted due to poor number statistics. \label{mf_fit}}
\end{figure*}

Figure \ref{mf_fit}  shows  the Schechter functions fitted to the mass functions of
galaxies in different environments for colour-selected galaxy samples 
(see Tables \ref{tab:fit_global} and \ref{tab:fit_col}). The fits are used
to extrapolate  toward lower masses. No additional normalisation has been applied.  
This figure clearly shows that in all environments blue galaxies dominate in number the mass functions at low masses. 
Hence, the shape of the total mass function at low masses 
is regulated by the shape of the blue mass functions.  Since the mass functions of blue galaxies have 
similar  $\alpha$ values, the similarity of the total mass functions in the different environments is explained (at least
for masses $<M_\ast^*$). In addition, the similarity of $M_\ast^*$ for red and blue galaxies in clusters, groups and 
the field makes the high mass end of the mass functions  similar in all environments.

\subsection{Global and local environment}\label{gl_loc}
In \cite{ld}, we analysed specifically the
dependence of the mass function on local density, using collectively the
WINGS, PM2GC, ICBS, and EDisCS samples. We found that both in the local
and distant universe, in clusters and in the field,
local density plays an important role in driving the mass
distribution. In general, lower density regions host
proportionally a larger number of  low-mass galaxies than higher density ones.
In the field,  local density regulates the shape of the
mass function for low and high galaxy masses.  
The situation in clusters is slightly different: local density
seems to be important only when we reach relatively low masses
($\log M_{\ast}/M_{\odot} \leq 10.1$ in WINGS and $\log M_{\ast}/M_{\odot} \leq
10.4$ in EDisCS). The local density seems to have no significant effect 
for more massive samples. 

In the same study we also found that not only the shape of the mass function is
different, but also the highest mass reached: the most massive
galaxies are located only the highest density regions, and they
are clearly absent in the the lowest densities studied (this is often called mass segregation).

Combining the results of \cite{ld} with those of this paper we 
find that galaxy samples with comparable mass selection limits 
have mass distributions that vary with local density but not with global environment.
In other words, global and local environment appear to influence the shape of the mass functions
in very different ways. While the global
environment seems to have little effect, local density is 
clearly important in determining
the most fundamental of all galaxy properties, its mass.
Clearly, the global and local environments are linked to different physical
processes. Their distinct effect on the mass function
needs  to be understood in order to understand the drivers of galaxy formation and transformation.

\subsection{Some caveats}
Obviously, the results presented in this paper are only valid for
galaxies in the mass ranges covered by our samples. 
We have no information for masses below $\log M_{\ast}/M_{\odot} \simeq 10.2$ in clusters and groups, and 
$\log M_{\ast}/M_{\odot} \simeq 10.5$ in the field.
At lower masses the situation could well be  very different, 
and deeper surveys are needed 
to establish the role of the environment for low-mass galaxies.

Nevertheless, the mass limits we have adopted are relatively low, so our
study is not only relevant to massive galaxies. 
For instance, the EDisCS mass limit ($M_{\ast}\sim 1.6\times 10^{10} M_{\odot}$) 
is lower than the critical mass proposed by \cite{kauffmann03}  
separating low-$z$ galaxies in two distinct populations 
with different stellar population ages, 
surface mass densities, concentrations, and star formation rates ($3 \times 10^{10} M_{\odot}$). 

The lack of evidence suggesting a 
dependence of the mass function on global environment 
could also be partly due to small number statistics, at least in some cases. 
Although most of our results seem reasonably reliable, in particular
those confirmed in both galaxy samples and those  with 
high statistical significance  (e.g., those derived for EDisCS), 
larger samples would still be desirable.
A larger spectroscopic 
sample of field galaxies would allow, for instance, a better assessment of 
whether truly isolated galaxies still follow the same mass distribution.  

Using photo-$z$ selection for EDisCS galaxies is also not ideal. 
The level of contamination, even if  reasonably low (see e.g. \citealt{halliday04, milvang08}), 
may conceivably still have some effect. 
Larger spectroscopic surveys would therefore help to confirm (or otherwise) our results.

Our results also depend on the adopted IMF. 
We implicitly assume that
it is universal, regardless of time, environment, and galaxy
morphological type. 
Of course, this may not always be the case. 
Reports of IMF variation with the star-formation rate
\citep{Guna11} or the velocity
dispersion of the galaxies \citep{treu09, vd11, thomas11, dutton11,
cappellari12, spiniello12} have recently appeared in the literature. 
Obviously, if the IMF is not
universal, our results may be quite different since the stellar mass
estimates could change in different ways at different galaxy masses. 
It is fair to say, though, that it would require quite a contrived variation 
of the IMF with galaxy mass in order to make intrinsically-different mass functions 
appear the same in all global environments. 

\section{Conclusions}\label{conc}

In this paper we have analysed the shape of the stellar  mass
function for galaxies in different environments: clusters, groups, and
the field.  We have used two mass-limited  samples covering different 
redshift ranges, ICBS ($0.3\leq z \leq 0.45$) and EDisCS 
($0.4\leq z \leq 0.8$).  The stellar mass ranges considered are  
$M_{\ast} \geq 10^{10.5} M_{\odot}$ for
ICBS and $M_{\ast} \geq 10^{10.2} M_{\odot}$ for EDisCS.
Our main results are:
\begin{itemize}
\item Galaxies in clusters  ($R\leq R_{200}$), groups, and the field 
seem to follow similar mass distributions. 
We find no statistical evidence suggesting a dependence of the shape of
the galaxy stellar mass function on global environment at $z=0.3$--$0.8$.

\item By comparing of the ICBS mass functions 
with mass functions in the local
universe (\citealt{morph} for cluster galaxies and  Calvi et al., in
preparation, for field ones), we find that 
the evolution  of the shape of the mass function from $z\sim0.4$ to $z\sim0.07$ is
the same in the field and in clusters, and hence independent
of the  global environment.
However, the amount of growth with time of the mass function is higher for low-mass galaxies than for massive ones.

\item Galaxies inhabiting the virialised regions of 
clusters at various clustercentric distances
have very similar mass functions. This is also the case for  
galaxies inside and outside $R_{200}$. 

\item Red and blue galaxies in clusters, groups and the field
have different mass functions. However, When comparing 
the mass functions separately for red and blue galaxies
in different environments, no differences are detected. 

\end{itemize}
 
In summary, we find that the global
environment does not seem to have a prominent role in shaping galaxy stellar mass functions.
In contrast, the local environment does seem to have significant influence in determining
the galaxy mass distribution \citep{ld}. This suggests 
that the most fundamental of all galaxy properties, its mass, is not very dependent 
on the mass of the halo it inhabits, but it does depend on local scale processes.
Global and local environments are clearly linked to different physical
processes. Understanding their distinct roles in altering galaxy properties
is important in order to understand the drivers of galaxy formation and evolution.

\begin{acknowledgements}
We thank the anonymous referee for her/his detailed and careful comments and suggestions which helped us to improve the manuscript. 
We thank Lucia Pozzetti for providing the $z$-COSMOS data for the field mass functions and useful discussion.
We also thank Micol Bolzonella for  her suggestions.
BV and BMP acknowledge financial support from ASI contract I/016/07/0 
and ASI-INAF I/009/10/0. BV also acknowledges financial support from the 
Fondazione Ing. Aldo Gini. GDL acknowledges financial support from the European Research Council under
the European Community's Seventh Framework Programme (FP7/2007-2013)/ERC
grant agreement n. 202781.
\end{acknowledgements}

\appendix
\section{Is the mass function simply a mirror of the luminosity function?}
Using exactly the same samples as in \S \ref{mf_envi}, in this Appendix we build 
the corresponding luminosity functions to test whether luminosity and mass functions 
yield equivalent information and results in terms of their environmental differences. 
We use the absolute 
magnitudes $M_V$ derived in \S \ref{dataicbs} and \S \ref{dataediscs} for both 
ICBS and EDisCS.  

In parallel with the mass functions, 
in each magnitude bin we sum all galaxies belonging to the
environment under consideration to obtain the total number of
galaxies. The width of
each magnitude bin corresponds to $0.4\,{\rm dex}$ in luminosity.
As for the mass functions, histograms are normalised  using the total integrated luminosity
down to the relevant limits.

\begin{figure*}
\centering
\includegraphics[scale=0.4]{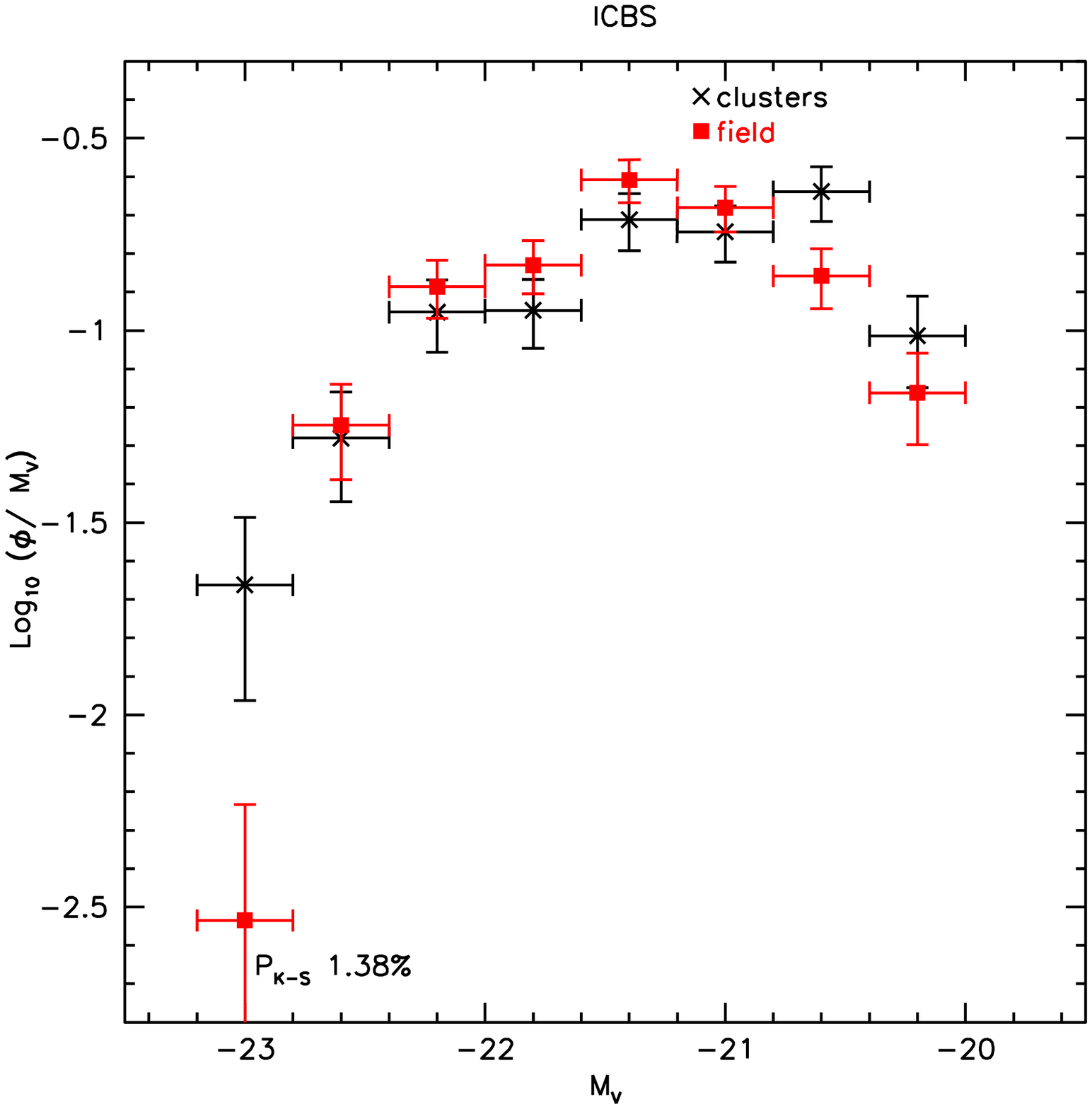}
\includegraphics[scale=0.4]{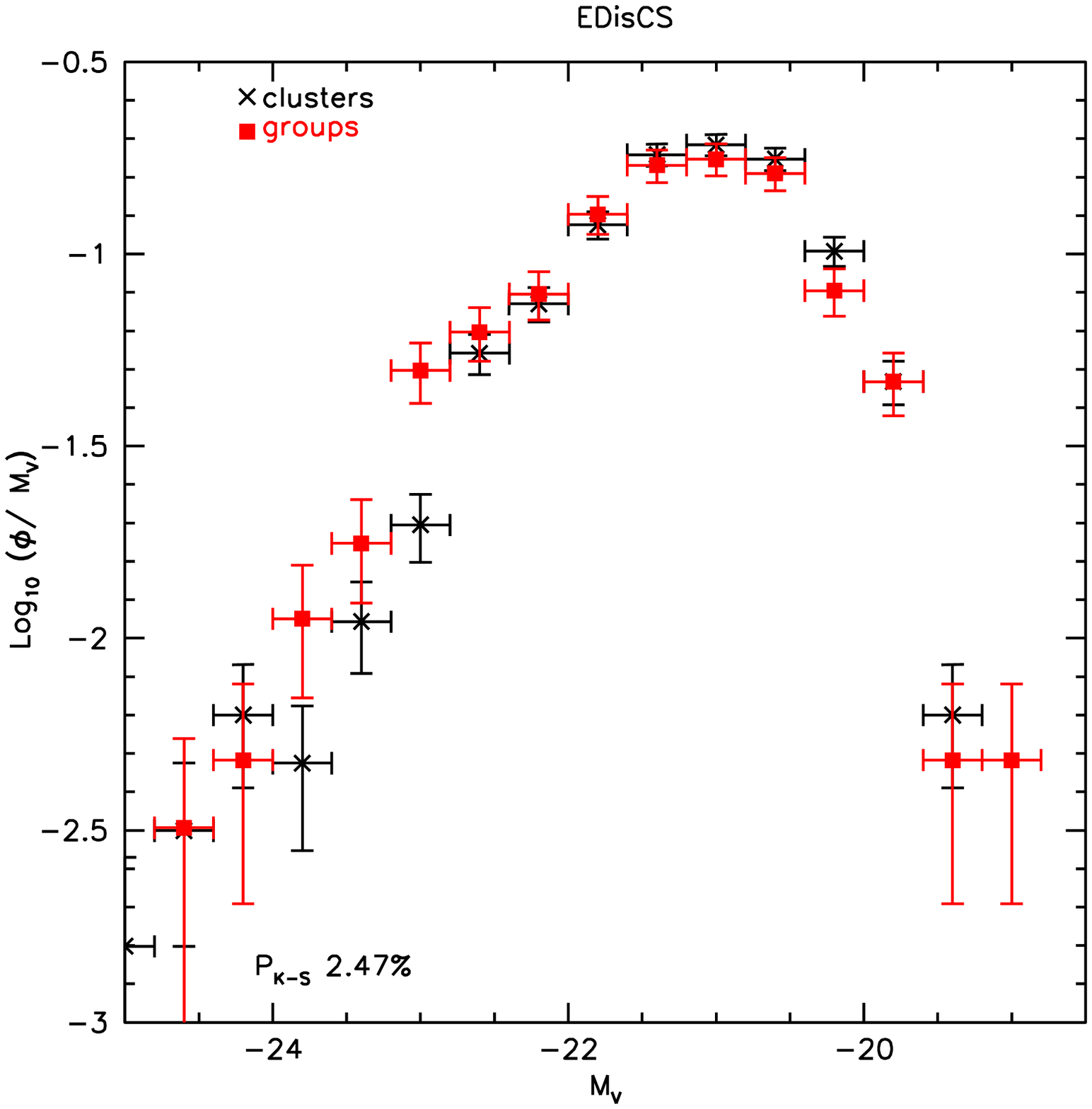}
\caption{Luminosity functions in different environments. Left panel: ICBS cluster regions (black crosses) and 
field (red filled squares). Right panel: EDisCS clusters (black crosses) and groups (red filled squares). 
In each panel, luminosity functions are normalised using the total integrated luminosity. At the bottom of each panel,
the K--S probabilities are given. Significant differences are evident between the luminosity functions in different environments. 
\label{lf}}
\end{figure*}

For the ICBS survey (left panel of \fig\ref{lf}), we compare cluster regions and the field. 
The two distributions are different: in clusters the
number of more luminous galaxies is proportionally higher than in the field.
A K--S test supports this visual impression giving a probability of $\sim 1.4\%$ for both
distributions being similar. 

In EDisCS  (right panel of \fig\ref{lf}), we compare the cluster regions with groups. Again, 
both from a visual inspection and from a K--S test ($P_{\rm K--S} \sim 2.5\%$), we conclude that the 
two distributions are different. Groups seem to have a higher number of more
luminous galaxies than clusters.

Therefore, using both ICBS and EDisCS data, the luminosity functions 
of galaxies in different global environments 
seem statistically different, while the mass functions are indistinguishable 
(see \S \ref{mf_envi}).
This shows that studying the luminosity function does not give direct
information on the mass function. Galaxies have different 
colours and mass-to-light ratios and therefore stellar  masses do not 
scale from luminosities by a constant factor.

\end{document}